\documentclass[lettersize,journal,onecolumn]{IEEEtran}
\usepackage{amsmath,amsfonts}
\usepackage{algorithmic}
\usepackage{array}
\usepackage[caption=false,font=normalsize,labelfont=sf,textfont=sf]{subfig}
\usepackage{textcomp}
\usepackage{stfloats}
\usepackage{url}
\usepackage{verbatim}
\usepackage{graphicx}
\def\BibTeX{{\rm B\kern-.05em{\sc i\kern-.025em b}\kern-.08em
    T\kern-.1667em\lower.7ex\hbox{E}\kern-.125emX}}
\usepackage{balance}

\usepackage{cancel}

\newcommand{\dd}{\textnormal{d}}

\usepackage{amssymb} % \triangleq
\usepackage[center]{caption} % to center the caption of the figure
\usepackage{tikz}
\usetikzlibrary {shapes.geometric} 
\usetikzlibrary{arrows.meta}
\allowdisplaybreaks
\usepackage[numbers, square, comma]{natbib}
\usepackage[amsmath]{ntheorem} % to put the theorem headlines in bold
\usepackage{dsfont}
\usepackage{bbm}

\usepackage{hyperref}
\usepackage{cleveref}
\hypersetup{
    colorlinks=true, %set true if you want colored links
    linktoc=all,     %set to all if you want both sections and subsections linked
    linkcolor=black,  %choose some color if you want links to stand out
    citecolor=black
}

\usepackage{setspace}
\newtheorem{theorem}{\textbf{Theorem}}

\newtheorem{definition}{\textbf{Definition}}
\newtheorem{remark}{\textbf{Remark}}

\newtheorem{example}{\textbf{Example}}
\newtheorem{proposition}{\textbf{Proposition}}

\newcommand{\Sn}{\boldsymbol{S}_n}
\newcommand{\sn}{\boldsymbol{s}}

\theoremstyle{plain}
\theoremheaderfont{\normalfont\itshape}
\theoremseparator{:}

\def\BibTeX{{\rm B\kern-.05em{\sc i\kern-.025em b}\kern-.08em
    T\kern-.1667em\lower.7ex\hbox{E}\kern-.125emX}}

\newcommand{\norm}[1]{\left\|#1\right\|}

\begin{document}
\title{Multi-Sensor Distributed Hypothesis Testing in the Low-Power Regime}
%\title{Distributed hypothesis testing over channels under a zero-rate communication constraint}
%\title{Distributed hypothesis testing with side information over additive memoryless channels}
%\title{Distributed hypothesis testing over additive memoryless channels}
\author{C\'{e}cile Bouette, Mich\`{e}le Wigger\\
\thanks{C\'{e}cile Bouette and Mich\`{e}le Wigger are with  LTCI, T\'{e}l\'{e}com Paris, Institut Polytechnique de Paris, France (email: cecile.bouette@telecom-paris.fr; michele.wigger@telecom-paris.fr).
This work was supported by the ERC under Grant Agreement 101125691.}}

%\markboth{Journal of \LaTeX\ Class Files,~Vol.~18, No.~9, September~2020}%
%{How to Use the IEEEtran \LaTeX \ Templates}

\maketitle

\begin{abstract}

We characterize the Stein-exponent of a distributed hypothesis testing scenario where two sensors transmit information through a memoryless multiple access channel (MAC) subject to a sublinear input cost constraint with respect to the number of channel uses and where the decision center has access to an additional local observation. Our main theorem provides conditions on the channel and cost functions for which the Stein-exponent of this distributed setup is no larger than the Stein-exponent of the local test at the decision center. Under these conditions, communication from the sensors to the decision center is thus useless in terms of Stein-exponent. The conditions are satisfied for additive noise MACs with generalized Gaussian noise under a $p$-th moment constraint (including the Gaussian channel with second-moment constraint) and for the class of fully-connected (where all inputs can induce all outputs) discrete memoryless multiple-access channels (DMMACs) under arbitrary cost constraints. We further show that for DMMACs that are not fully-connected, the Stein-exponent is larger and coincides with that of a setup with zero-rate noiseless communication links from either both sensors or only one sensor, as studied in \cite{papamarcou92}.
\end{abstract}


\section{Introduction}

Binary hypothesis testing refers to a problem that involves determining which of two joint distributions governs observed data. This is a standard problem encountered in many sensor applications and, as such, also in sensor networks and the Internet of Things (IoT). A specificity of the IoT is that sensors have extremely stringent power budgets because their batteries are supposed to last for decades. Recent 6G standards tighten the requirement on the sensor's power consumptions even further, in particular under the framework of Ambient IoT \cite{3gpp_release}. Our goal in this paper is to study the performance of distributed binary hypothesis testing under stringent power budgets at the sensors. In particular, we will impose stringent power constraints on the signals that are transmitted by the sensors. 

Formally, a sensor network consists of several sensors, all with local observations, and at least one decision center that is tasked to decide on one of two hypotheses $H\in\{0,1\}$ based on the information that is communicated from the sensors and possibly also based on own local observations. The goal of the sensors and the decision center is to minimize the two types of error: the type-I error probability which refers to the probability that the decision center declares $\hat{H}=1$ while the true hypothesis underlying the observations is $H=0$; and the type-II error probability which refers to the probability that the decision center declares $\hat{H}=0$ while the true hypothesis underlying the observations is $H=1$. We are interested in asymmetric situations where type-II error probabilities are more harmful than type-I error probabilities because hypothesis $H=0$ describes a normal situation, while $H=1$ describes an alert situation such as a tsunami or avalanche event.  In such scenarios, type-I errors are often referred as false-alarm events and type-II errors as miss-detection events. To capture the asymmetry in the hypothesis test, the Stein-exponent  \cite{stein1972bound} measures the largest possible decay rate to zero (in the number of observations $n$) of the type-II error probability under a fixed threshold $\epsilon \in [0,1)$ on the type-I error probability.

The Stein-exponent is well-known for local tests where all the observations are locally available at the decision center  \cite{stein1972bound}, in which case it does not depend on the allowed type-I error threshold $\epsilon \in[0,1)$. For most distributed scenarios where part of the observations are located at remote sensors and first need to be communicated to the decision center, the Stein-exponent is however still unknown. Notable exceptions are, for example, \cite{ papamarcou92,hypothesis_testing_ahlswede,hypothesis_testing_te_han, sreejith_hypothesis_noisy_dmc, variable_length_hypothesis_wigger, optimality_binning_hypothesis,sreekumar_without_side_information,  escamilla2020distributed}, which make different assumptions on the communication from the sensors to the decision center.

A canonical line of work  \cite{hypothesis_testing_ahlswede,hypothesis_testing_te_han, optimality_binning_hypothesis, Kochman, yuvalligong, SHA,Amin,Watanabe_GHT, zhao2018distributed,Kochman-MAC} studied the Stein-exponent in a communication scenario where the single sensor can send $R n$ bits to the decision center over a noise- and error-free link. For a small class of source distributions, the Stein-exponent has been determined, but it remains open in general and only upper and lower bounds are available. Extensions were also proposed for multi-sensors networks \cite{optimality_binning_hypothesis,zhao2018distributed,Kochman-MAC,wang} or under a variable-length coding framework \cite{variable_length_hypothesis_wigger,Multihop_Hamad, ITW20}.

A line of work that is more closely related to the present work studied the Stein-exponent again in a noiseless-link setup, but under the constraint that the number of bits communicated from the sensor to the decision center grows only sublinearly in $n$. This setting is commonly known as \emph{noiseless zero-rate communication}, and \cite{papamarcou92,hypothesis_testing_te_han} characterized the Stein-exponent for a broad class of source distributions. In particular, Han~\cite{hypothesis_testing_te_han} considered the single-sensor setup where the sensor can only send a single bit to the decision center. The optimal strategy in this scenario is for the decision center to decide on $\hat{H}=0$ if, and only if, both its own observation and the sensor's observation are \emph{marginally typical} under the distribution corresponding to hypothesis $H=0$. In this strategy, the sensor only sends the binary outcome of its typicality test to the decision center.
Shalaby and Papamarcou \cite{papamarcou92} proved that this simple 1-bit communication strategy and the corresponding decision rule are optimal and achieve the Stein-exponent even in scenarios where the sensor can send a sublinear number of bits to the decision center. They further also extended the result to multi-sensor setups \cite{papamarcou92}, and more recently, similar zero-rate results have been obtained in the quantum domain~\cite{Sreekumar}.

In this work we will relax the assumption of noise-free communication and instead consider general classes of memoryless multiple-access channels (MAC) from the sensors to the decision center. In the information-theoretic literature, noisy communication channels for hypothesis testing have been explored in~\cite{sreejith_hypothesis_noisy_dmc, Michele_noisy_and_MAC}, presenting general lower bounds and exact characterizations of the Stein-exponent for special source distributions.  In these works, channel input sequences have same length as the observations and no cost constraints are imposed. 

A recent work \cite{itw_dichotomy} also considered distributed hypothesis testing over noisy communication links,  but under the assumption that the input sequences either have to be much shorter than the observations or they are subject to stringent block-input cost constraints that grow only sublinearly in the number of observations. These assumptions are motivated by stringent resource constraints at the sensors, as mentioned in the first paragraph. More specifically, the work in \cite{itw_dichotomy} focuses on a single-sensor setup where communication is over a discrete-memoryless channel (DMC). It is shown that under the two mentioned input constraints (sublinear number of inputs or cost), the Stein-exponent only depends on whether the DMC is fully-connected, i.e.,   each input symbol induces each output symbol with positive probability, or not. For fully-connected DMCs, the Stein-exponent completely degrades to the Stein-exponent of the local test at the decision center rendering the sensor and its communication useless. For partially-connected DMCs, in contrast,  the Stein-exponent is equal to the  Stein-exponent of a setup where communication from the sensor to the decision center takes place over a zero-rate \emph{noiseless} link as studied in \cite{papamarcou92}. It was thus shown that imposing stringent constraints on the input sequence strongly degrades the Stein-exponent over DMCs, and for certain channels even renders communication useless.

In this paper, we extend the results in \cite{itw_dichotomy} to scenarios with multiple sensors that can communicate to the decision center over a continuous or discrete memoryless MAC. We assume that the inputs from each of the two sensors are subject to stringent block-input cost constraints that grow only sublinearly in the number of observations $n$. For a large class of MACs, we show a complete degradation of the Stein-exponent in this setup to the Stein-exponent of a local test, rendering the communication and the sensors useless. As we  further show, this is in particular the case for 
\begin{itemize} 
\item continuous-valued  MACs with $p$-th order generalized Gaussian noise subject to a  $p$-th moment block-input constraint; and
\item fully-connected DMMACs, i.e., DMMACs where each input pair induces each output pair with positive probability,  with arbitrary nonnegative block-input constraints.
\end{itemize} 
For other connectivity patterns of the DMMAC, the Stein-exponent over DMMACs coincides with the Stein-exponent of a communication scenario where either one or both sensors can communicate to the decision center over zero-rate noiseless links, as in \cite{papamarcou92}.

To summarize, in this paper we show that for a multi-sensor distributed hypothesis testing setup the Stein-exponent can completely collapse to the local Stein-exponent when stringent sublinear cost-constraints are imposed on the channel input sequences. Exceptions are channels where some outputs can only be induced by certain input pairs, where the degradation can be milder and Stein-exponents as in noiseless zero-rate communication scenarios remain achievable. Here, again, the type of communication scenario (one or two noiseless links) to consider depends on the structure (connectivity) of the original MAC. Our results thus extend the previous results in \cite{itw_dichotomy} in two directions: multiple sensors and continuous channels. Both extensions cannot be obtained from the results in \cite{itw_dichotomy} but required new proof elements. 
\medskip

The rest of this paper is organized as follows. Section \ref{section_notation} introduces  notation  and Section \ref{section_setup}
describes the general problem setup. Section \ref{sec:results} presents our main results, which state the following.
\begin{itemize} 
\item Conditions for continuous and discrete memoryless MACs and associated cost-constraints are presented under which the Stein-exponent degrades to the local Stein-exponent at the decision center.
\item The additive noise MAC with generalized Gaussian noise of parameter $p>0$ under a $p$-th moment cost constraints satisfies above conditions, and thus its  Stein-exponent degrades to the local Stein-exponent.
\item The Stein-exponent for the general class of DMMACs under arbitrary cost constraints. Its Stein-exponent depends on the connectivity of the DMMAC, ranging from the local Stein-exponent to the Stein-exponent of a zero-rate noiseless link scenario from either both or only a single sensor.  
\end{itemize}
The subsequent Sections~\ref{sec:proof1} and \ref{sec:proof2} present the proofs of our main results, and Section \ref{section_conclusion}  presents concluding remarks. 

\section{Notation}
\label{section_notation}

Random variables are denoted by uppercase letters like $U$, while their realizations are denoted by lowercase letters like $u$. We abbreviate $(u_1,\ldots, u_n)$  by  $u^n$ and the subvector $(u_i,\ldots, u_j),i\leq j,$ by  $u^j_i$. Depending on the context, if a vector $u_l$ is itself indexed by $l$, its subvector $(u_{l,i},\ldots,u_{l,j}),i\leq j,$ is denoted by $u_{l,i}^j$. The set of all real numbers is denoted $\mathbb{R}$ and the set of nonnegative real numbers by $\mathbb{R}^+$. We use $P_U$ to denote the law (also called distribution) of the random variable $U$ and $P_{U^n}$ that of the random vector $U^n$.
The associated Lebesgue-Stieltjes  measure \cite[Section 3.11]{williams_probability_2018} of $P_U$ is denoted by $\dd P_U$. When it exists, the probability density function corresponding to the distribution $P_U$ is denoted $p_U$. We denote the product of measures by $\otimes$ and the distribution of an independent and identically distributed sequence of random variables $U^n$ by $P_{U}^{\otimes n}$. 
We abbreviate \emph{independent and identically distributed} as \emph{i.i.d.}. The set of all probability distributions over the set $\mathcal{U}$ is denoted by $\mathcal{M}(\mathcal{U})$.
%and \textcolor{red}{\emph{probability mass function} as \emph{pmf} : à enlever ?  ne parler que de distribution ?}.
%The set of all probability distributions over the set $\mathcal{U}$
%the measurable space $(\mathcal{U},\mathcal{T}(\mathcal{U}))$, where $\mathcal{T}(\mathcal{U})$ is a $\sigma$-algebra on $\mathcal{U}$,
%is denoted by $\mathcal{M}(\mathcal{U})$.
 Given a sequence $u^n\in\mathcal{U}^n$,
%\textcolor{red}{for $\mathcal{U}$ a finite set},
we denote its type \cite[Ch. 11]{coverthomas06} by
\begin{equation}
\pi_{u^n}(a)\triangleq\frac{\left|\{i\in\{1,\ldots,n\}:u_i=a\}\right|}{n}, \qquad a\in \mathcal{U}.
\end{equation}
The set of all possible $n$-types is denoted by $\mathcal{P}_n(\mathcal{U})=\{\pi_{u^n}:u^n\in\mathcal{U}^n\}$.
We denote the type class \cite[Ch. 11]{coverthomas06} of $\pi_{u^n}$ as $\mathcal{T}_n(\pi_{u^n})=\{\tilde{u}^n\in\mathcal{U}^n: \pi_{\tilde{u}^n}=\pi_{u^n}\}$.
For $\mu>0$,  the strongly-typical set \cite[Ch. 10]{coverthomas06} for a given a random variable $U$ is denoted by $\mathcal{T}_{\mu}(P_{U})$ and is defined as the set of all sequences $u^n\in\mathcal{U}^n$ with type $\pi_{u^n}$ satisfying $|\pi_{u^n}(a)-P_U(a)|\leq \mu$, $\forall a\in\mathcal{U}$, and $\pi_{u^n}(a)=0$ if $P_U(a)=0$.
Likewise,
%for two finite sets $\mathcal{U}$ and $\mathcal{V}$
we denote the joint type of the sequences $(u^n, v^n)\in\mathcal{U}^n\times\mathcal{V}^n$ by
\begin{equation}
\pi_{u^n,v^n}(a,b)\triangleq\frac{\left|\{i\in\{1,\ldots,n\}:u_i=a,v_i=b\}\right|}{n}, \quad (a,b)\in  \mathcal{U}\times \mathcal{V}.
\end{equation}
%The class of equivalence for type $\pi_{u^n,v^n}$ is $\mathcal{T}_n(\pi_{u^n,v^n})=\{\tilde{u}^n,\tilde{v}^n\in\mathcal{U}^n\times\mathcal{V}^n, \pi_{\tilde{u}^n:\tilde{v}^n}=\pi_{u^n,v^n}\}$. %We define the jointly strongly-typical set $\mathcal{T}_{\mu}(P_{U,V})$, $\mu>0$, as the set of all pair of sequences $(u^n,v^n)$ with respective types $\pi_{u^n},\pi_{v^n}$ satisfying  $|\pi_{u^n}(a)-P_U(a)|\leq \mu$ and $|\pi_{v^n}(a)-P_V(a)|\leq \mu$, $\forall a\in\mathcal{U}$.
Let $P_1$ and $P_2$ be two distributions over $\mathcal{U}$. Suppose that $P_1$ is absolutely continuous with respect to $P_2$ (denoted $P_1 \ll P_2$) which means that for any $u\in\mathcal{U}$, $P_2(u)=0\implies P_1(u)=0$, then the Kullback-Leibler divergence
%\cite{kullback_leibler,reverse_pinsker}
between $P_1$ and $P_2$ is denoted
\begin{IEEEeqnarray}{rCl}
\label{def_kl_divergence}
    {D}(P_1\|P_2)&=&\int_{\mathcal{U}}\ln\left(\frac{\dd P_1}{\dd P_2}\right) \dd P_1,
\end{IEEEeqnarray}
where $\frac{\dd P_1}{\dd P_2}$ is the Radon–Nikodym  derivative \cite[p. 128]{halmos2013measure} of $P_1$ with respect to $P_2$.
%We denote by $Q$ the Gaussian tail function\index{Gaussian tail function}:
%%definition $\varepsilon$ see Theorem 28
%\begin{IEEEeqnarray}{rCl}
%Q:~&&\mathbb{R}\rightarrow\mathbb{R}^+\nonumber\\
%~&&x\mapsto\int_x^{\infty} \frac{1}{\sqrt{2\pi}}e^{-\frac{t^2}{2}} \dd t.
%\end{IEEEeqnarray}
We denote the quasi $p$-norms by $\norm{a^n}_p=\left(\sum_{i=1}^n |a_i|^p\right)^{1/p}$ for any $a^n\in\mathbb{R}^n$. %Finally,  $\Gamma$ denotes the gamma function \cite{gamma_and_digamma_functions}.

\section{General Model}
\label{section_setup}

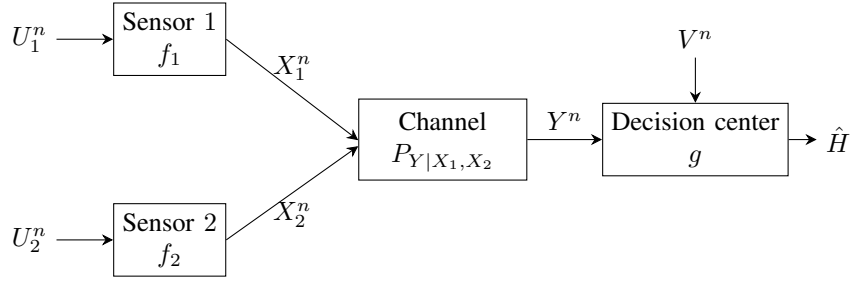
\begin{figure}[!htbp]
\begin{center}
        \begin{tikzpicture}[
            nodetype1/.style={
                rectangle,
                %rounded corners,
                minimum width=0.7cm,
                minimum height=0.7cm,
                draw=black,
                font=\normalsize
            },
            nodetype2/.style={
                rectangle,
                %rounded corners,
                minimum width=0.55cm,
                minimum height=0.7cm,
                draw=black,
                font=\normalsize
            },
            tip2/.style={-{Stealth[length=1.5mm, width=1.5mm]}},
            every text node part/.style={align=center}
            ]
            \matrix[row sep=0.3cm, column sep=0.38cm, ampersand replacement=\&]{
            %\& \& (invisible) \&\& (invisible) \& \& (invisible) \& \& \& \node (V) {$V^n$}; \\
            \node (U_1) {$U_1^n$}; \&\& \node (encoder1) [nodetype2]   {Sensor $1$\\$f_1$};
             \& \& (invisible) \& \& \& (invisible) \& \node (V) {$V^n$};\&; \\
            %\& \& (invisible) \& \& (invisible) \&\& \\
            \&  \& (invisible) \& \& (invisible) \& \& 
            \node (W)  [draw, nodetype1, text width=2cm, text centered]  {\begin{tabular}{c} Channel\\$P_{Y|X_1,X_2}$ \end{tabular}}; \&
            \node (Y){};
            \&
            \node (decoder) [nodetype2] {Decision center\\$g$}; \&
            \node (guess) {$\hat{H}$};\\
            \node (U_2) {$U_2^n$};  \& \& \node (encoder2) [nodetype2]  {Sensor $2$\\$f_2$}; \&
            \node (X){}; \\
            };
            
            \draw[->] (U_1) edge[tip2] node [above] {} (encoder1) ;
            \draw[-]  (encoder1.east) edge[tip2] node [above] (X1) {$X_1^n$} (W.mid west) ;
            \draw[->] (W) edge[tip2] node [above] {$Y^n$} (decoder) ;
            \draw[->] (decoder) edge[tip2] node [above] {} (guess) ;
            \draw[-] (encoder2.east) edge[tip2] node [below] (X2) {$X_2^n$} (W.base west) ;
            \draw[->] (U_2) edge[tip2] (encoder2) ;
            \draw[->] (V) edge[tip2] (decoder) ;
        \end{tikzpicture}
    \caption{Distributed hypothesis testing over a discrete memoryless channel.}
     \label{fig:multi-users_hypothesis_testing}
     \end{center}
\end{figure}

We consider the setup of Figure \ref{fig:multi-users_hypothesis_testing} where two sensors observe random sequences $U_1^n$ and $U_2^n$, respectively, and a decision center observes the random sequence $V^n$. The observations take value in finite sets $\mathcal{U}_1^n, \mathcal{U}_2^n$, and $\mathcal{V}^n$, respectively, and their distribution depends on a binary hypothesis $H\in\{0,1\}$ in the sense that:
\begin{subequations}
\label{definition_hypothesis}
\begin{IEEEeqnarray}{rCl}
  \textnormal{under }  H = 0 &\colon& \quad (U_1^n, U_2^n, V^n)\sim P_{U_1,U_2,V}^{\otimes n}\\
  \textnormal{under }    H = 1&\colon&\quad  (U_1^n, U_2^n, V^n) \sim Q_{U_1,U_2,V}^{\otimes n},
\end{IEEEeqnarray}
\end{subequations}
where we assume that $P_{U_1,U_2,V}\ll Q_{U_1,U_2,V}$.

The two sensors communicate over a memoryless multiple-access channel (MAC) with a decision center, which then attempts to guess the hypothesis $H$ based on its own local observations and on its observed channel outputs. More formally, each Sensor~$\ell\in \{1,2\}$ produces a sequence of channel inputs $X_\ell^n=f_\ell(U^n_\ell)$ using an encoding function of the form
\begin{IEEEeqnarray}{rCl}
f_\ell \colon \mathcal{U}_\ell^n & \rightarrow& \mathcal{X}_\ell^n \nonumber\\
 u_\ell^n& \mapsto &x_\ell^n = f_\ell(u_\ell^n), \quad \ell \in \{1,2\}.
\end{IEEEeqnarray}
The input sequences are subject to stringent cost constraints. That means,  we are given per-symbol input cost functions 
\begin{IEEEeqnarray}{rCl}c_\ell \colon \mathcal{X}_\ell & \rightarrow&\mathbb{R}^+, %\\
% u_\ell^n& \mapsto &x_\ell^n = f_\ell(u_\ell^n), \quad \ell \in \{1,2\}.
\end{IEEEeqnarray}
where we assume that there exists a unique zero-symbol $x_\ell \in \mathcal{X}_\ell$ with $c_\ell(x_\ell)=0$. For simplicity, we call  0  the zero-symbol for both $c_1$ and $c_2$.
We are also given  stringent cost budgets  $\Gamma_1(n)>0$ and $\Gamma_2(n)$ that grow sublinearly in $n$: 
\begin{IEEEeqnarray}{rCl}\label{eq:stringent}
  \lim_{n\to \infty} \Gamma_\ell(n) =\infty \qquad \textnormal{ and } \qquad   \lim_{n\to \infty} \frac{\Gamma_\ell(n)}{n}=0,\quad \ell \in\{1,2\}.
\end{IEEEeqnarray}
For each $u_\ell^n$, the cost constraints impose that $f_\ell(u_\ell^n)$ lies in
\begin{equation} 
\tilde{\mathcal{X}}^n_\ell \triangleq\left \{ x_\ell^n \in \mathcal{X}_\ell^n \colon   \sum_{i=1}^n c_l(x_{\ell,i}) \leq \Gamma_\ell(n) \right\}, \quad  \ell \in\{1,2\}. \label{cost_constraint}
\end{equation} 
Given  channel inputs $x_1^n$ and $x_2^n$, the decision center observes a random output sequence $Y^n$ with each $Y_i$ distributed according to  the conditional distribution  $P_{Y|X_1,X_2}(\cdot|x_1,x_2)$. 

The decision center applies a guessing function of the form
\begin{IEEEeqnarray}{rCl} g \colon \mathcal{Y}^n \times  \mathcal{V}^n & \rightarrow& \{0,1\} %\\
\end{IEEEeqnarray}
to produce a random guess $\hat{H}=g(Y^n,V^n)$. 
The decision's   type-I (false positive) and type-II (false negative) error probabilities are respectively denoted
\begin{subequations}
\begin{IEEEeqnarray}{rCl}
    \alpha_n&=&\mathbb{P}\left[ g(Y^n,V^n)=1\; \big| \;H=0\right]\\
    \beta_n&=&\mathbb{P}\left[g(Y^n,V^n)=0\; \big| \;H=1\right].
\end{IEEEeqnarray}
\end{subequations}
\begin{definition}
Given  $\epsilon\in[0,1)$, a type-II error exponent $\theta>0$ is called $(\epsilon, \Gamma_1(\cdot), \Gamma_{2}(\cdot))$-achievable  if there exists a  sequence (in the blocklength $n$) of encoding and decision functions   $(f_1,f_2,g)$  satisfying the sublinear cost constraint \eqref{cost_constraint} and
\begin{subequations}
\label{error_exponent_definition}
\begin{IEEEeqnarray}{rCl}
	\varlimsup_{n\to \infty}  \alpha_n &\leq&  \epsilon \\
    \varliminf_{n\to \infty} - \frac{1}{n} \ln\beta_n &\geq&  \theta.
\end{IEEEeqnarray} 
\end{subequations}
The supremum over all $(\epsilon,\Gamma_1(\cdot),\Gamma_2(\cdot))$-achievable  type-II  error exponents $\theta$ is denoted $\theta_{\textnormal{sublin}}^*(\epsilon,\Gamma_1(\cdot), \Gamma_2(\cdot))$.
 \end{definition}
We shall see that $\theta_{\textnormal{sublin}}^*(\epsilon,\Gamma_1(\cdot),\Gamma_2(\cdot))$ is the same  for all functions $\Gamma_1(\cdot)$  and $\Gamma_2(\cdot)$ satisfying \eqref{eq:stringent} and all $\epsilon \in [0,1)$. We therefore simply write $\theta^*_{\textnormal{sublin}}$.

The following proposition will be instrumental in the results of this paper. The proof can be found in Appendix~\ref{app:Randomness}.
\begin{proposition}\label{lemma2}
Allowing the decision center to take a randomized decision does not increase the Stein-exponent $\theta^*_{\textnormal{sublin}}$. 
\end{proposition}

\section{Results}\label{sec:results}

We have the following general result on the type-II error exponent showing that for certain channels $P_{Y|X_1,X_2}$, communication from the sensors does not increase the Stein-exponent due to the stringent cost constraints  \eqref{eq:stringent}. 

\begin{theorem}
\label{thm1}
Assume that the channel $P_{Y|X_1,X_2}$ is such that there exists a sequence of sets $\mathcal{D}_n \subseteq \mathcal{Y}^n$ satisfying
\begin{equation} \label{eq:Dn}
\lim_{n\to \infty}\mathbb{P}[ Y^n \notin \mathcal{D}_n]=0
\end{equation} and for all input sequences   $x_1^n, \tilde{x}_1^n \in \tilde{ \mathcal{X}}_1^n$ and  $x_2^n,\tilde{x}_2^n \in \tilde{ \mathcal{X}}_2^n$:
\begin{equation} \label{eq:ratio_lim} 
\varlimsup_{n \to \infty} - \frac{1}{n} \ln \left(\frac{\dd P_{Y|X_1,X_2}^{\otimes n} (y^n|x_1^n,x_2^n)}{\dd P_{Y|X_1,X_2}^{\otimes n} (y^n|\tilde{x}_1^n,\tilde{x}_2^n)}\right) \leq  0, \quad \forall y^n\in \mathcal{D}_n.
\end{equation} 
Then, for any $\epsilon \in [0,1)$ and $\Gamma_1(\cdot)$ and $\Gamma_2(\cdot)$ satisfying \eqref{eq:stringent}, the optimal type-II error exponent is    
\begin{IEEEeqnarray}{rCl}
        \theta_{\textnormal{sublin}}^* = D\left(P_{V} \| Q_{V} \right).
\end{IEEEeqnarray}
\end{theorem}
\begin{IEEEproof}
See Section~\ref{sec:proof_thm1}.
\end{IEEEproof}

\begin{remark}
   The optimal exponent $\theta_{\textnormal{sublin}}^*$ of Theorem \ref{thm1} coincides with the optimal exponent in a scenario where the decision center only observes the local observation $V^n$ \cite[Theorem 14.13]{info_theory_polyanskiy}. Therefore, under the stringent cost constraints \eqref{eq:stringent}, the sensors cannot improve the Stein-exponent.
\end{remark}

\begin{remark}
The same Stein-exponent holds also in the related setup where the sensors communicate to the decision center only over $k(n)$ channel uses where $k(n)$ grows sublinearly in $n$,  irrespective of whether a block-input power constraint or a per-symbol power constraint is imposed. This holds because achievability of the Stein-exponent does not need any communication and the converse is implied by the converse above, as the setup is weaker. In fact, in our original setup, the sensor can always choose to only transmit during the first $k(n)$ channel uses.
\end{remark}

\subsection{Generalized Gaussian channels with a $p$-th moment constraint}
\label{section_gen_gaussian}

Consider the  MAC
\begin{IEEEeqnarray}{rCl}
\label{gg_channel}
 Y_i=  h_1 x_{1,i} + h_2 x_{2,i} + Z_i,  \qquad  i\in\{1,\ldots,n\},
\end{IEEEeqnarray}
where $h_1$ and $h_2$ are given non-zero real channel coefficients, $(Z_i)_{i\geq 1}$ is an  i.i.d. sequence independent of the inputs $x_1^n$ and $x_2^n$,  and each $Z_i$ follows a generalized Gaussian distribution \cite{nadarajah2005,generalized_gaussian_distributions_short} with parameters $p,\sigma>0$, i..e, of probability density function 
\begin{IEEEeqnarray}{rCl}
    p_Z(z)=\frac{c_p}{\sigma}e^{-\frac{|z|^p}{2\sigma^p}}, \qquad z\in\mathbb{R}, 
\end{IEEEeqnarray}
where
\begin{IEEEeqnarray}{rCl}
 c_p \triangleq \frac{p}{2^{\frac{p+1}{p}}\Gamma\left(\frac{1}{p}\right)}
 \end{IEEEeqnarray}
and $\Gamma$ denotes the gamma function \cite{gamma_and_digamma_functions}.
 A $p$-th moment  cost constraint is imposed on the input sequences: 
\begin{IEEEeqnarray}{rCl}
    \label{eq:ppower}
\| x_{\ell}^n\|_p^p\leq \Gamma_\ell(n), \quad \ell \in\{1,2\},
\end{IEEEeqnarray}
i.e., $c_\ell(x)=|x|^p, \ell=1,2$.

For  $p=2$, the noise is Gaussian and the cost constraint is a standard average block-power constraint.

\begin{theorem}
\label{cor2}
For the above  generalized Gaussian setup, for any $p >0$,  $\epsilon \in [0,1)$, and $p$-th moment cost constraints $\Gamma_1(\cdot)$ and  $\Gamma_2(\cdot)$ satisfying \eqref{eq:stringent}:
    \begin{IEEEeqnarray}{rCl}
        \theta_{\textnormal{sublin}}^* = D\left(P_{V} \| Q_{V} \right).
\end{IEEEeqnarray}
\end{theorem}

\begin{IEEEproof}
See Section~\ref{sec:proof_cor2}.
\end{IEEEproof}

\subsection{Discrete memoryless channels with arbitrary cost constraints}
\label{section_dmc}

Communication takes place over a discrete memoryless MAC (DMMAC). Accordingly, input and output sets $\mathcal{X}_1, \mathcal{X}_2$, and $\mathcal{Y}$ are finite and the channel law is described by a probability mass function (pmf) $P_{Y|X_1,X_2}$. As we shall see, $   \theta_{\textnormal{sublin}}^*$ depends on the topology (connectivity) of the DMMAC. We therefore define the following classes of channels. 
\begin{itemize} 
\item The set $\mathcal{C}_{\textnormal{full}}$ contains all DMMACs $P_{Y|X_1,X_2}$ for which  
\begin{IEEEeqnarray}{rCl}
    P_{Y|X_1,X_2}(y|x_1, x_2) &>& 0, \qquad \forall y,x_1,x_2 \in\mathcal{Y}\times\mathcal{X}_1\times\mathcal{X}_2. 
\end{IEEEeqnarray}
\item The set $\mathcal{C}_{\textnormal{sparse}}$ contains all DMMACs $P_{Y|X_1,X_2}$ where there exists (not necessarily distinct) $x_1, x_1', \tilde{x}_1\in\mathcal{X}_1$, $x_2, x_2',\tilde{x}_2\in\mathcal{X}_2$ and $y,\tilde{y}\in\mathcal{Y}$ so that:
\begin{subequations}\label{eq:full_conditions}
\begin{IEEEeqnarray}{rCl}
    P_{Y|X_1,X_2}(y|x_1, \tilde{x}_2) &=& 0 \\
    P_{Y|X_1,X_2}(y|x_1', \tilde{x}_2) &>& 0 \\
    P_{Y|X_1,X_2}(\tilde{y}|\tilde{x}_1, x_2) &=& 0  \\
    P_{Y|X_1,X_2}(\tilde{y}|\tilde{x}_1, x_2') &>& 0.
\end{IEEEeqnarray}
\end{subequations}
\item The set $\mathcal{C}_{\textnormal{sparse-full}}$ contains  all DMMACs $P_{Y|X_1,X_2}$ for which
there exists $x_1,x_1'\in \mathcal{X}_1$, $\tilde{x}_2\in\mathcal{X}_2$ and $y^*\in \mathcal{Y}$ so that
\begin{subequations}
\begin{IEEEeqnarray}{rCl}
\label{definition_C_sparse_full}
    P_{Y|X_1,X_2}(y^*|x_1, \tilde{x}_2) &=&0\\
    P_{Y|X_1,X_2}(y^*|x_1', \tilde{x}_2)&>&0,
\end{IEEEeqnarray}
\end{subequations}
and  for any pair $(x_1,y)\in\mathcal{X}_1\times\mathcal{Y}$ we  either have
\begin{subequations}
\begin{IEEEeqnarray}{rCl}
    P_{Y|X_1,X_2}(y|x_1, x_2)&=&0, \quad\forall x_2\in\mathcal{X}_2
    \end{IEEEeqnarray}
    or 
  \begin{IEEEeqnarray}{rCl}  
    P_{Y|X_1,X_2}(y|x_1, x_2)&>&0, \quad \forall x_2\in\mathcal{X}_2.
\end{IEEEeqnarray}
\end{subequations}
\item  Finally, the set $\mathcal{C}_{\textnormal{full-sparse}}$ contains all DMMACs $P_{Y|X_1,X_2}$ for which 
there exists $\tilde{x}_1\in \mathcal{X}_1$, $x_2, x_2'\in\mathcal{X}_2$ and $y^*\in\mathcal{Y}$ so that
\begin{subequations}
\begin{IEEEeqnarray}{rCl}
    P_{Y|X_1,X_2}(y^*|\tilde{x}_1, x_2) &=&0\\
    P_{Y|X_1,X_2}(y^*|\tilde{x}_1, x_2')&>&0,
\end{IEEEeqnarray}
\end{subequations}
and  for any pair $(x_2,y)\in\mathcal{X}_2\times\mathcal{Y}$ we  either have
\begin{subequations}
\begin{IEEEeqnarray}{rCl}
    P_{Y|X_1,X_2}(y|x_1, x_2)&=&0, \quad\forall x_1\in\mathcal{X}_1
    \end{IEEEeqnarray}
    or 
  \begin{IEEEeqnarray}{rCl}  
    P_{Y|X_1,X_2}(y|x_1, x_2)&>&0, \quad \forall x_1\in\mathcal{X}_1.
\end{IEEEeqnarray}
\end{subequations}
\end{itemize}
To see that the four sets $\mathcal{C}_{\textnormal{full}},\mathcal{C}_{\textnormal{sparse}},\mathcal{C}_{\textnormal{sparse-full}},\mathcal{C}_{\textnormal{full-sparse}}$ are a partition of all channel laws, notice that the event
  \begin{IEEEeqnarray}{rCl} 
  \{ \exists x_1,\tilde{x}_1, y^* \colon  \quad P_{Y|X_1,X_2}(y^*|x_1, \tilde{x}_2)=0 \qquad{\textnormal{and}} \qquad P_{Y|X_1,X_2}(y^*|x_1', \tilde{x}_2)>0\}
  \end{IEEEeqnarray}
  is the complementary event of 
\begin{IEEEeqnarray}{rCl} 
\{ \forall x_1  \colon\; P_{Y|X_1,X_2}(y^*|x_1, \tilde{x}_2)=0 \} \quad \cup \quad    \{\forall x_1  \colon\; P_{Y|X_1,X_2}(y^*|x_1, \tilde{x}_2)>0 \} .
\end{IEEEeqnarray}
\medskip

\begin{example}
Consider the DMMAC
\begin{equation}
Y= S_1 \cdot x_1 +S_2 \cdot x_2 + Z,
\end{equation}
where multiplications and additions are the standard operations in $\mathbb{R}$ and $Z$ is a discrete noise over any discrete and finite set $\mathcal{Z}\subset \mathbb{R}$. Inputs $x_1$ and $x_2$ as well as the auxiliary variables $S_1$ and $S_2$ take value in $\{-1,1\}$. Assuming that $Z$ is not deterministic, depending on the law of the ``states" $S_1$ and $S_2$, the DMC belongs to one of the four classes above. In fact,  
\begin{itemize}
\item For $S_1$ and $S_2$ both deterministic, the DMMAC belongs to $\mathcal{C}_{\textnormal{sparse}}$.
\item For $S_1$ and $S_2$ both non-deterministic,  the DMMAC belongs to $\mathcal{C}_{\textnormal{full}}$.
\item For $S_1$ deterministic and $S_2$  non-deterministic,  the DMMAC belongs to $\mathcal{C}_{\textnormal{sparse-full}}$.
\item For $S_1$ non-deterministic and $S_2$  deterministic,  the DMMAC belongs to $\mathcal{C}_{\textnormal{full-sparse}}$.
\end{itemize}
\end{example}
      \medskip
\begin{theorem}
\label{theorem_dmc}
%Consider the hypothesis testing problem with side information \eqref{definition_hypothesis} over a DMC with cost constraint \eqref{cost_constraint}.
\begin{enumerate}
\item \label{it:th1} If $P_{Y|X_1,X_2}\in \mathcal{C}_{\textnormal{full}}$,
then
\begin{IEEEeqnarray}{rCl}
       \theta_{\textnormal{sublin}}^* = D\left(P_{V} \| Q_{V} \right).
\end{IEEEeqnarray}
\item \label{it:th2} If  $P_{Y|X_1,X_2}\in\mathcal{C}_{\textnormal{sparse}}$, 
then
\begin{IEEEeqnarray}{rCl}
    \theta_{\textnormal{sublin}}^* = \min D\left( \tilde{P}_{U_1,U_2,V} \| Q_{U_1,U_2,V} \right),
\end{IEEEeqnarray}
where the minimum is taken over all distributions $\tilde{P}_{U_1,U_2,V}$ satisfying
\begin{IEEEeqnarray}{rCl}
\tilde{P}_{U_1} = P_{U_1}, \quad \tilde{P}_{U_2} = P_{U_2}, \quad \tilde{P}_{V} = P_{V}.    
\end{IEEEeqnarray}%If $\exists x_1, x_1', \tilde{x}_1, x_2, x_2', y', y'$ such that
\item \label{it:th3} If $P_{Y|X_1,X_2}\in\mathcal{C}_{\textnormal{sparse-full}}$, then
\begin{IEEEeqnarray}{rCl}
    \theta_{\textnormal{sublin}}^* = \min D\left( \tilde{P}_{V,U_1} \| Q_{V,U_1} \right),
\end{IEEEeqnarray}
where the minimum is taken over all distributions $\tilde{P}_{V,U_1}$ such that
\begin{IEEEeqnarray}{rCl}
\tilde{P}_{U_1} = P_{U_1}, \quad \tilde{P}_{V} = P_{V}.    
\end{IEEEeqnarray}
\item \label{it:th4} If $P_{Y|X_1,X_2}\in \mathcal{C}_{\textnormal{full-sparse}}$, then
\begin{IEEEeqnarray}{rCl} 
    \theta_{\textnormal{sublin}}^* = \min D\left( \tilde{P}_{V,U_2} \| Q_{V,U_2} \right),
\end{IEEEeqnarray}
where the minimum is taken over all distributions $\tilde{P}_{V,U_2}$ such that
\begin{IEEEeqnarray}{rCl}
\tilde{P}_{U_2} = P_{U_2}, \quad \tilde{P}_{V} = P_{V}.    
\end{IEEEeqnarray}
\end{enumerate} 
%irrespectively of $0<\epsilon<1$.
\end{theorem}
\begin{IEEEproof}
See Section~\ref{sec:proof2}.
\end{IEEEproof}
\begin{remark} 
In case \ref{it:th1}) the optimal exponent is the same as in a scenario without any sensor or without communication from the sensors to the decision center. In case \ref{it:th2}), the optimal exponent is the same as when both sensors can communicate to the decision center over independent zero-rate noiseless links. In case \ref{it:th3}), the optimal exponent is the same as in a scenario without  Sensor $2$ and where Sensor 1 communicates to the decision center over a zero-rate noiseless link, see   \cite[Theorem 1]{papamarcou92}.
\end{remark}

\begin{remark}
Our results in Theorem~\ref{theorem_dmc} remain valid when the DMMAC can only be used for a sublinear (in $n$) number of times $k(n)$, irrespective of whether an input-cost constraint is imposed or not. The converse results follow in a straightforward way because this new setup is weaker (in our original setup, the sensors can always choose to transmit only during the first $k(n)$ channel uses). Inspecting the achievability proofs of Theorem~\ref{theorem_dmc}, we see that communication from each sensor effectively only takes place over a sublinear number of channel uses at the beginning, while during the rest of the communication, both sensors send the all-zero sequence. These latter channel inputs can thus be omitted without any loss of information at the decision center. The proposed Stein-exponents in Theorem~\ref{theorem_dmc} can thus also be achieved in our new setup where communication is only over $k(n)$ channel uses.
\end{remark}

\section{Proof of Theorem \ref{thm1} }\label{sec:proof_thm1}\label{sec:proof1}

Achievability follows directly from Stein's lemma, where the decision center can ignore the channel outputs $Y^n$. 
The converse is proved by relating the type-I and type-II error probabilities of our distributed hypothesis testing problem to the error probabilities of a randomized local hypothesis testing setup. %and secondly bounding the probabilities of this related setup. 
We start our proof by fixing any sequence of  encoding and decision functions $\{f_1, f_2,g\}_{n=1}^\infty$ with type-I error satisfying
\begin{equation}
\varlimsup_{n\to \infty} \alpha_n\leq\epsilon.
\end{equation}

For each observation $v^n\in\mathcal{V}^n$, define the acceptance regions \begin{IEEEeqnarray}{rCl}
\mathcal{A}(v^n) \triangleq \{ y^n \in \mathbb{R}^n \colon g( v^n, y^n)=0\}.
\end{IEEEeqnarray}
%and the set of inputs
%\begin{IEEEeqnarray}{rCl}
%\tilde{\mathcal{X}}_\ell^n \triangleq \left \{ x_\ell^n \in \mathbb{R}^n \colon x_\ell^n= f_\ell(u_\ell^n) \textnormal{ for some }u_\ell^n\in\mathcal{U}_\ell^n \right\}.
%\end{IEEEeqnarray}

%\noindent\textit{Step 1: The related randomized local hypothesis testing problem:} 

\begin{figure}[!htbp]
\begin{center}
        \begin{tikzpicture}[
            nodetype1/.style={
                rectangle,
                %rounded corners,
                minimum width=0.7cm,
                minimum height=0.7cm,
                draw=black,
                font=\normalsize
            },
            nodetype2/.style={
                rectangle,
                %rounded corners,
                minimum width=0.55cm,
                minimum height=0.7cm,
                draw=black,
                font=\normalsize
            },
            tip2/.style={-{Stealth[length=1.5mm, width=1.5mm]}}
            ]
            \matrix[row sep=0.3cm, column sep=0.38cm, ampersand replacement=\&]{
            \& \& \node (V) {$V^n$}; \\
            \& \& \node (invisible) {}; \& \\
            \node (W)  [draw, nodetype1, text width=2cm, text centered]  {$P_{\tilde{Y}^n|V^n}$}; \&
            \node (Y_tilde){};
            \&
            \node (decoder) [nodetype2] {$\tilde{g}$}; \&
            \node (guess) {$\hat{H}$};\\
            };
            
            \draw[arrows = {-Latex[length=1pt]}] (W) edge[tip2] node [above] {$\tilde{Y}^n$} (decoder) ;
            \draw[->] (decoder) edge[tip2] node [above] {} (guess) ;
            \draw[-{Stealth[length=1.5mm, width=1.5mm]}] (invisible.center) -| (W) ;
            \draw[-{Stealth[length=1.5mm, width=1.5mm]}] (invisible.center) -- (decoder) ;
            \draw[-] (V) -- (invisible.center) ;
        \end{tikzpicture}
\caption{Randomized local hypothesis test.}

\label{fig:2}
\end{center}
\end{figure}
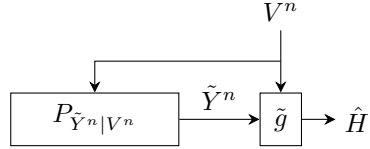

The chosen functions $f_1, f_2,g$  imply a  joint distribution on $U^n,V^n,X_1^n, X_2^n,Y^n$ under both hypotheses $H=0$ and $H=1$. We are particularly interested in the induced conditional probability
distribution
\begin{equation} \label{eq:cond}
P_{\tilde{Y}^n|V^n}(y^n|v^n)\triangleq \mathbb{P}[ Y^n = y^n| V^n=v^n, H=0]
\end{equation} 
and introduce the  new binary hypothesis testing setup depicted in Figure~\ref{fig:2} where under ${H}=0$  the decision center observes $V^n$ i.i.d. $\sim P_V$ and under ${H}=1$ it observes $V^n$ i.i.d. $\sim Q_V$. Moreover, under both hypotheses, it has access to an additional local randomness $\tilde{Y}^n$ that is obtained from $V^n$ based on the non-i.i.d. conditional distribution $P_{\tilde{Y}^n|V^n}$, irrespectively of the hypothesis $H$. The decision function in this auxiliary setup is thus of the form $\tilde{g}\colon \mathcal{V}^n\times \mathcal{Y}^n \to \{0,1\}$ and we  denote its type-I and type-II error probabilities by $\tilde \alpha_n$ and $\tilde \beta_n$: 
\begin{subequations} 
\begin{IEEEeqnarray}{rCl}
\tilde \alpha_n & \triangleq& \mathbb{P}[ \tilde g(V^n,\tilde{Y}^n)=1 | H=0]  \\
\tilde \beta_n & \triangleq  &  \mathbb{P}[ \tilde g(V^n,\tilde{Y}^n)=0 | H=1]  .
\end{IEEEeqnarray}
\end{subequations}

Choosing  the decision rule
\begin{equation} \label{eq:gg}
\tilde{g}(v^n,y^n) = g(v^n,y^n) \cdot \mathbbm{1}\{ y^n \in \mathcal{D}_{\delta,n}\}
\end{equation}
for  the setup  of Figure~\ref{fig:2}, 
we have:
  \begin{IEEEeqnarray}{rCl}
    1-\alpha_n&=&  \mathbb{P}\left[ g( V^n, Y^n) =0 \; \big| \; H=0\right] \\
    &=&  \mathbb{P}\left[ g( V^n, Y^n) =0, Y^n \in \mathcal{D}_{n}  \; \big| \; H=0\right] + \mathbb{P}\left[ g ( V^n, Y^n) =0, Y^n \notin \mathcal{D}_{n}  \; \big| \; H=0 \right] \\  
    &\leq&  \mathbb{P}\left[ g( V^n, Y^n) =0, Y^n \in \mathcal{D}_{n}  \; \big| \; H=0\right] + \mathbb{P}\left[  Y^n \notin \mathcal{D}_{n}  \; \big| \; H=0\right]  \\
    &=&\sum_{v^n\in\mathcal{V}^n} \Big( \mathbb{P}\left[ g( V^n, Y^n) =0, Y^n \in \mathcal{D}_{n}  \; \big| \; V^n=v^n, H=0\right]P^{\otimes n}_{V}(v^n) \Big) + \mathbb{P}\left[  Y^n \notin \mathcal{D}_{n}  \; \big| \; H=0\right] \label{eq:proof_lemma_1_alpha_step_1a0}  \\
    &=&\sum_{v^n\in\mathcal{V}^n}\left( \mathbb{P}\left[ g( V^n, \tilde{Y}^n) =0, \tilde{Y}^n \in \mathcal{D}_{n}  \; \big| \; V^n=v^n, H=0\right] P^{\otimes n}_{V}(v^n) \right)+ \mathbb{P}\left[  Y^n \notin \mathcal{D}_{n}  \; \big| \; H=0\right]     \label{eq:proof_lemma_1_alpha_step_1a} \\
    &=& \mathbb{P}\left[ \tilde{g}( V^n, \tilde{Y}^n) =0 \; \big| \; H=0\right] + \mathbb{P}\left[  Y^n \notin \mathcal{D}_{n}  \; \big| \; H=0\right]  \\
    \label{eq:proof_lemma_1_alpha_step_final}
    &=& 1- \tilde{\alpha}_n + \mathbb{P}\left[  Y^n \notin \mathcal{D}_{n}  \; \big| \; H=0\right],
 \end{IEEEeqnarray}
where \eqref{eq:proof_lemma_1_alpha_step_1a0} holds because we defined $\tilde{Y}^n$ to have the same conditional distribution given $V^n$ as $Y^n$ under hypothesis $H=0$ and \eqref{eq:proof_lemma_1_alpha_step_1a} holds by the definition of the $\tilde{g}$ function in \eqref{eq:gg}.
Combining above inequality with \eqref{eq:Dn}, we deduce that 
\begin{equation}\label{eq:abound}
\varlimsup_{n \to \infty} \tilde{\alpha}_n\leq  \varlimsup_{n \to \infty}  \alpha_n \leq \epsilon.
\end{equation}
Define
\begin{equation} \label{eq:zeta}
\zeta \triangleq \inf_{ \substack{y^n \in \mathcal{D}_n,\\
x_1^n, \tilde{x}_1^n \in \tilde{\mathcal{X}}_1^n,\\
x_2^n, \tilde{x}_2^n \in \tilde{\mathcal{X}}_2^n}} \frac{ \dd P_{Y|X_1,X_2}^{\otimes n} (y^n |x_1^n,x_2^n)}{  \dd P_{Y|X_1,X_2}^{\otimes n} (y^n |\tilde{x}_1^n, \tilde{x}_2^n) }.
\end{equation}  
We can now bound $\beta_n$ in terms of $\tilde{\beta}_n$: 
\begin{IEEEeqnarray}{rCl}
    \beta_n
    &=&\sum_{v^n\in\mathcal{V}^n}\mathbb{P}[Y^n\in\mathcal{A}(v^n),V^n=v^n|H=1]\\
      &=&\sum_{v^n\in\mathcal{V}^n}\left(Q_{V}^{\otimes n}(v^n)\int_{\mathcal{A}(v^n)} \dd P_{Y^n|V^n,H}(y^n|v^n,1) \right)\\
    &\geq &\sum_{v^n\in\mathcal{V}^n}\left(Q_{V}^{\otimes n}(v^n)\int_{\mathcal{A}(v^n)\cap \mathcal{D}_n} \sum_{(x_1^n,x_2^n)\in \tilde{\mathcal{X}}_1^n\times \tilde{\mathcal{X}}_2^n} \left(P_{X_1^n,X_2^n|V^n,H}(x_1^n,x_2^n|v^n,1) \cdot \dd P_{Y|X_1,X_2}^{\otimes n}(y^n|x_1^n,x_2^n) \right) \right) \label{eq:1}\\
    &\geq& \zeta  \sum_{v^n\in\mathcal{V}^n}\left(Q_{V}^{\otimes n}(v^n) \int_{\mathcal{A}(v^n) \cap \mathcal{D}_n}\sum_{(x_1^n,x_2^n)\in\tilde{\mathcal{X}}_1^n\times\tilde{\mathcal{X}}_2^n} \left(  P_{X_1^n,X_2^n|V^n,H}(x_1^n,x_2^n|v^n,0)  \cdot  \dd P_{Y|X_1,X_2}^{\otimes n}(y^n|x_1^n,x_2^n) \right)\right)     \label{eq:proof_lemma_1_beta_step_1a}\\
    &=&\zeta  \sum_{v^n\in\mathcal{V}^n}\left(Q_{V}^{\otimes n}(v^n)\int_{\mathcal{A}(v^n) \cap \mathcal{D}_{n}}  \dd P_{\tilde{Y}^n|V^n}(y^n|v^n)\right)    \label{eq:proof_lemma_1_beta_step_2a}\\
& = & \zeta  \cdot \tilde{\beta}_n,   \label{identifying_beta_tilde0}
\end{IEEEeqnarray}
where \eqref{eq:1} holds by restricting the integral and by the total law of probability; 
\eqref{eq:proof_lemma_1_beta_step_1a} holds by the definition of $\zeta$ in \eqref{eq:zeta} and because for any bounded function $f$\footnote{Notice that by Assumption~\eqref{eq:ratio_lim}, for any $y^n\in\mathcal{D}_n$, the function $(x_1^n,x_2^n)\mapsto \dd P_{Y|X_1,X_2}^{\otimes n} (y^n |{x}_1^n, {x}_2^n)$ is bounded with finite non-zero infimum and supremum.} the expectations of $f$ with respect to any two measures $\mu$ and $\nu$ satisfy $\mathbb{E}_{\nu}[f] / \mathbb{E}_{\mu}[f] \geq \frac{f_{\min}}{f_{\max}}$ where $f_{\min}$ and $f_{\max}$  denote the infimum and supremum of $f$;
%\eqref{eq:proof_lemma_1_beta_step_1a} \textcolor{blue}{follows by noticing that for any $y^n\in\mathcal{D}_n$,}
%\textcolor{blue}{
%\begin{IEEEeqnarray}{rCl}
%\IEEEeqnarraymulticol{3}{l}{
%\sum_{(x_1^n,x_2^n)\in\tilde{\mathcal{X}}^n_1\times \tilde{\mathcal{X}}^n_2} P_{X_1^n,X_2^n|V^n,H}(x^n|v^n,1) \cdot \dd P_{Y|X_1,X_2}^{\otimes n}(y^n|x^n_1, x^n_2)
%}\nonumber\\* ~~~~~~~~~~~~~~~~~~~~~~
%&\geq&  \inf\limits_{(x_1^n,x_2^n)\in\tilde{\mathcal{X}}^n_1\times \tilde{\mathcal{X}}^n_2} \dd P_{Y|X_1,X_2}^{\otimes n}(y^n|x^n_1, x^n_2)\\
%&\geq& \sum_{(x_1^n,x_2^n)\in\tilde{\mathcal{X}}^n_1\times \tilde{\mathcal{X}}^n_2} P_{X_1^n,X_2^n|V^n,H}(x^n|v^n,0) \inf\limits_{(x_1^n,x_2^n)\in\tilde{\mathcal{X}}^n_1\times \tilde{\mathcal{X}}^n_2} \dd P_{Y|X_1,X_2}^{\otimes n}(y^n|x^n_1, x^n_2) \nonumber\\
%&&{} ~~~~~~~~~~~~~~~~~~~~~~\times\frac{\dd P_{Y|X_1,X_2}^{\otimes n}(y^n|x^n_1, x^n_2)}{\sup\limits_{(x_1^n,x_2^n)\in\tilde{\mathcal{X}}^n_1\times \tilde{\mathcal{X}}^n_2}\dd P_{Y|X_1,X_2}^{\otimes n}(y^n|x^n_1, x^n_2)},
     %&\geq& \sum_{(x_1^n,x_2^n)\in\tilde{\mathcal{X}}^n_1\times \tilde{\mathcal{X}}^n_2} P_{X_1^n,X_2^n|V^n,H}(x^n|v^n,0) \dd P_{Y|X_1,X_2}^{\otimes n}(y^n|x^n_1, x^n_2)\xi;
%\end{IEEEeqnarray}
%}
%\textcolor{blue}{and by the definition of $\zeta$ in \eqref{eq:zeta}}; 
\eqref{eq:proof_lemma_1_beta_step_2a} holds by the definition of $\tilde{Y}^n$ in \eqref{eq:cond}; \eqref{identifying_beta_tilde0} holds by the definition of $\tilde{\beta}_n$.\\

Notice now that we can specialize Proposition~\ref{lemma2} to a setup with a useless MAC from the two sensors to the decision center, in which case the decision center can base its decision only on the local observation $V^n$ and the local randomness. Applying this proposition to the setup of Figure~\ref{fig:2} where the local randomness is $\tilde{Y}^n$, we  conclude that $\frac{1}{n} \ln \tilde{\beta}_n$ is asymptotically upper bounded by the Stein-exponent of a non-random test based on $V^n$ only, i.e., 
\begin{equation} 
\varlimsup_{n\to \infty} - \frac{1}{n} \ln\tilde \beta_n \leq D(P_V \| Q_V). \label{eq:S}
\end{equation} 
Plugging  \eqref{eq:S} and Assumption~\eqref{eq:ratio_lim}   into  \eqref{identifying_beta_tilde0}, we obtain the desired converse result.

\section{Proofs of Theorem \ref{cor2}}
\label{sec:proof_cor2}
We only prove the converses since achievabilities are obvious. 

Start by noticing:
\begin{itemize}
    \item For $p\in(0,1]$ and for all $a,b\in\mathbb{R}$, 
    \begin{IEEEeqnarray}{rCl}
    \label{inequality_p_leq_1_bis}
    |a+b|^p\leq (|a|+|b|)^p\leq|a|^p+|b|^p,    
    \end{IEEEeqnarray}
    where the second inequality is proved in \cite[Eq (2.12.2)]{hardy1952inequalities}. Above inequalities also imply:
    \begin{IEEEeqnarray}{rCl}
    \label{inequality_p_leq_1}
    ||a|^p-|b|^p|\leq|a-b|^p.
    \end{IEEEeqnarray}
    \item For $p\in(1,\infty)$ and for all $a,b\in\mathbb{R}$, it holds that:
    \begin{IEEEeqnarray}{rCl}
    \label{inequality_p_geq_1}
    |a+b|^p \leq ( |a|+|b| )^p\leq 2^{p-1} (|a|^p+|b|^p),
    \end{IEEEeqnarray}
    where the first inequality holds by the triangle inequality and the second by the convexity of the $t\mapsto |t|^p$ function.
\end{itemize}
In particular, we  have for any $p>0$: 
\begin{equation} \label{eq:pgen}
  |a+b|^p \leq 2^{p} (|a|^p+|b|^p).
\end{equation} 
For ease of notation, we define
\begin{equation} 
b_i \triangleq h_1 x_{1,i} + h_2 x_{2,i}, \quad i\in\{1,\ldots, n\},
\end{equation}
and notice that  \eqref{eq:pgen} and the  input power constraints \eqref{eq:ppower} imply  that
\begin{equation} \label{eq:pb}
\| b^n \|_p^p \leq 2^{p} (h_1^p\Gamma_1(n)+h_2^p\Gamma_2(n)),
\end{equation} 
for any $x_1^n\in\tilde{\mathcal{X}}^n_1$ and $x_2^n\in \tilde{\mathcal{X}}^n_2$.

\medskip
\textit{Case $p\in(0,1]$}: The converse follows directly from Theorem \ref{thm1} by choosing $\mathcal{D}_n= \mathcal{Y}^n$. In fact, %\textcolor{blue}{recalling the cost constraint \eqref{eq:stringent}}, 
for any input sequences $x_1^n, \tilde{x}_1^n\in  \tilde{\mathcal{X}}^n_1$ and $x_2^n, \tilde{x}_2^n\in \tilde{\mathcal{X}}^n_2$, define $b^n= h_1x_1^n +h_2x_2^n$ and $\tilde{b}^n= h_1\tilde{x}_1^n +h_2\tilde{x}_2^n$, and notice:
\begin{IEEEeqnarray}{rCl}
\frac{ p_{Y|X_1,X_2}^{\otimes n}(y^n |x_1^n,x_2^n)}{ p_{Y|X_1,X_2}^{\otimes n}(y^n |\tilde{x}_1^n,\tilde{x}_2^n)}
%&=&\frac{e^{-\frac{\norm{y^n-x_1^n-x_2^n}_p^p}{2\sigma^p}}}{e^{-\frac{\norm{y^n-\tilde{x}_1^n - \tilde{x}_2^n}_p^p}{2 \sigma^p}}}\nonumber\\
&=&\exp\left( -\frac{\norm{ y^n- b^n }_p^p -\|y^n- \tilde{b}^n\|_p^p}{2\sigma^p }\right) \label{eq:ratio0}  \\
    &\geq  & \exp\left( -\frac{\|b^n - \tilde{b}^n\|_p^p}{2\sigma^p }\right)     \label{eq:ratio_1a}   \\
    &\geq&\exp\left(- \frac{\norm{b^n }_p^p + \|\tilde{b}^n\|_p^p }{2\sigma^p}\right)     \label{eq:ratio_2a}\\
    %&\leq&\exp\left(\frac{\sqrt{8(\Gamma_1(n)+\Gamma_2(n))} }{\sigma^2}\norm{y^n}_2 \right) \cdot\exp\left(\frac{\Gamma_1(n)+\Gamma_2(n)}{\sigma^2}\right)\\
    &\geq  & \exp\left(-\frac{ 2^{p} (h_1^p\Gamma_1(n)+h_2^p\Gamma_2(n))}{\sigma^p}\right) \label{eq:ratio_3a}
\end{IEEEeqnarray}
for any sequence $y^n \in \mathbb{R}^n$. 
Here, \eqref{eq:ratio_1a} holds  by \eqref{inequality_p_leq_1}; \eqref{eq:ratio_2a} holds by \eqref{inequality_p_leq_1_bis}; and \eqref{eq:ratio_3a} holds by \eqref{eq:pb}.

The proof is concluded by noting that \eqref{eq:ratio_3a} implies \eqref{eq:ratio_lim} because $p, h_1,h_2, \sigma$ are fixed and by our assumption \eqref{eq:stringent}.
\medskip

\textit{Case $p>1$}:  Follows  from Theorem \ref{thm1} by choosing 
\begin{IEEEeqnarray}{rCl}
\label{definition_D_delta_randomized_lemma_generalized_gaussian}
    \mathcal{D}_{n}&=&\left\{y^n\in\mathbb{R}^n:\norm{y^n}^{p}_p\leq \nu\right\},
\end{IEEEeqnarray}
where
\begin{IEEEeqnarray}{rCl}
\label{choice_nu_gg}
\nu&\triangleq &2^{2p-2} \left(h_1^p\Gamma_1(n)+h_2^p\Gamma_2(n)\right)+2^{p-1}\left(n\frac{2\sigma^p}{p}+ \delta n\right)
\end{IEEEeqnarray}
for an arbitrary small number $\delta>0$.

We check that the proposed set $\mathcal{D}_n$ satisfies the two Conditions \eqref{eq:Dn} and \eqref{eq:ratio_lim}   in the theorem.  For any $x_1^n, \tilde{x}_1^n\in\tilde{\mathcal{X}}_1^n$ and $x_2^n, \tilde{x}_2^n\in\tilde{\mathcal{X}}_2^n$, define $b^n$ and $\tilde{b}^n$ as above and notice:
\begin{IEEEeqnarray}{rCl}
\frac{ p_{Y|X_1,X_2}^{\otimes n}(y^n |x_1^n,x_2^n)}{ p_{Y|X_1,X_2}^{\otimes n}(y^n |\tilde{x}_1^n,\tilde{x}_2^n)}
%&=&\frac{e^{-\frac{\norm{y^n-x_1^n-x_2^n}_p^p}{2\sigma^p}}}{e^{-\frac{\norm{y^n-\tilde{x}_1^n - \tilde{x}_2^n}_p^p}{2 \sigma^p}}}\nonumber\\
&=&\exp\left( -\frac{\norm{ y^n- b^n }_p^p -\|y^n- \tilde{b}^n\|_p^p}{2\sigma^p }\right) \\
\label{eq_use_mean_theorem}
    &\geq&\exp\left(-\sum_{i=1}^n \frac{p|b_{i}  -\tilde{b}_{i} |\sup_{t\in(y_i-b_{i},y_i-\tilde{b}_i)}|t|^{p-1}}{2\sigma^p}\right)\\
    & \geq &  \exp\left(-\sum_{i=1}^n \frac{p\left(|b_{i}|  +|\tilde{b}_{i} | \right)\cdot {\max\left( (|y_i|+|b_i|)^{p-1}, (|y_i|+|\tilde{b}_i|)^{p-1}\right)}}{2\sigma^p}\right) \label{eq:sup}\\
  & \geq &  \exp\left(-\sum_{i=1}^n \frac{2^{p-1}p\left(|b_{i}|  +|\tilde{b}_{i} | \right)\cdot {\max\left( (|y_i|^{p-1}+|b_i|^{p-1}, |y_i|^{p-1}+|\tilde{b}_i|^{p-1}\right)}}{2\sigma^p}\right) \label{eq_gg_second_case:ratio_2}\\
    &\geq&\exp\Bigg(-\frac{2^{p-1}p}{2\sigma^p}\sum_{i=1}^n \Big(\left(|b_i| + |\tilde{b}_{i}|\right) \left(|y_i|^{p-1}+|b_i|^{p-1}+|\tilde{b}_i|^{p-1}\right)\Big)\Bigg) \\
    \label{eq_gg_second_case:ratio_3}
    %&\leq&\exp\Bigg(\frac{p}{2\sigma^p}\Bigg(\norm{x_{1}^n}_p^p + \norm{x_{2}^n}_p^p +\norm{x_{1}^n}_p\left(\sum_{i=1}^n |y_i|^p\right)^{\frac{p-1}{p}} + \norm{x_{2}^n}_p\left(\sum_{i=1}^n |y_i|^p\right)^{\frac{p-1}{p}} \nonumber\\
    %&&~~~~~~~~~~~~~~~~+ \norm{x_{1}^n}_p\left(\sum_{i=1}^n |x_{2,i}|^p\right)^{\frac{p-1}{p}} + \norm{x_{2}^n}_p\left(\sum_{i=1}^n |x_{1,i}|^p\right)^{\frac{p-1}{p}}\Bigg)\Bigg)\\
    &\geq&\exp\Bigg(-\frac{2^{p-2}p}{\sigma^p}\Big(\norm{b^n}_p^p + \|\tilde{b}^n\|_p^p  + \left(\norm{b^n}_p+\|\tilde{b}^n\|_p\right)\norm{y^n}_p^{p-1}\nonumber\\
    &&\hspace{3cm} +  \norm{b^n}_p\|\tilde{b}^n\|_p^{p-1}  + \|\tilde{b}^n\|_p\norm{b^n}_p^{p-1}\Big)\Bigg)\\
    \label{use_constraint_gg_p_geq_1}
    &\geq&\exp\left(-\frac{2^{p-2} p}{\sigma^p} \left(4 \cdot 2^{p} (h_1^p\Gamma_1(n)+h_2^p\Gamma_2(n)) + 2\left(h_1^p\Gamma_1(n)+h_2^p\Gamma_2(n) \right)^{\frac{1}{p}} \nu^{\frac{p-1}{p}} \right) \right),
\end{IEEEeqnarray}
where \eqref{eq_use_mean_theorem} holds by the mean value theorem and because the derivative of $|t|^p$ is $p |t|^{p-1}\cdot \textnormal{sign}(t)$; \eqref{eq:sup} holds because the supremum  is achieved at the borders of the interval;  \eqref{eq_gg_second_case:ratio_2} holds by \eqref{eq:pgen}; 
\eqref{eq_gg_second_case:ratio_3} holds by  factoring out and applying Hölder's inequality with the parameters $p$ and $\frac{p}{p-1}$ to the cross terms; and finally 
\eqref{use_constraint_gg_p_geq_1} holds by \eqref{eq:pb} and   the definition of $\mathcal{D}_n$ in \eqref{definition_D_delta_randomized_lemma_generalized_gaussian}.

Recalling the choice of $\nu$ in \eqref{choice_nu_gg} and the stringent power constraints \eqref{eq:stringent}, we can conclude from  \eqref{use_constraint_gg_p_geq_1} that Condition~\eqref{eq:ratio_lim}  in the theorem is satisfied.

To verify the  remaining Condition \eqref{eq:Dn}, define $B^n= h_1X_1^n+h_2X_2^n$ and notice the following:
\begin{IEEEeqnarray}{rCl}\label{eq:ineq}
\|Y^n \|_p^p & \leq & 2^{p-1}(\| B^n\|_p^p + \|Z^n\|_p^p) \label{in1} \\
& \leq & 2^{2p-2} \left(h_1^p\Gamma_1(n)+h_2^p\Gamma_2(n)\right)  + 2^{p-1} \|Z^n\|_p^p,\label{in2}
\end{IEEEeqnarray}
where \eqref{in1} holds by \eqref{inequality_p_geq_1}
%\eqref{use_inequality_p_geq_1}  
and \eqref{in2} by the power constraint \eqref{eq:ppower} and again \eqref{inequality_p_geq_1}. 
We thus have 
\begin{IEEEeqnarray}{rCl}
\mathbb{P}\left[Y^n\notin\mathcal{D}_{n}\; |\; H=0\right]&=&\mathbb{P}\left[\norm{Y^n}_p^p\geq \nu  \;\big | \; H=0\right]
    \label{use_inequality_p_geq_1}\\
%    &\leq&\mathbb{P}\left[\norm{X^n}^p_p+\norm{Z^n}^p_p\geq (\Gamma_1(n)+\Gamma_2(n))2^{p-1}+\left(\frac{2\sigma^p}{p}n+n\right)\right]\\
%    \label{use_2_inequality_p_geq_1}
%    %&\leq&\mathbb{P}\left[2^{p-1}\left(\Gamma_1(n)+\Gamma_2(n)\right)+\norm{Z^n}^p_p\geq (\Gamma_1(n)+\Gamma_2(n))2^{p-1}+\left(\frac{2\sigma^p}{p}n+\delta n\right)\right]\\
    &\leq&\mathbb{P}\left[ \frac{1}{n} \norm{Z^n}^p_p\geq  \left(\frac{2\sigma^p}{p}+ \delta n\right)\right]\\
    &=&\mathbb{P}\left[\frac{1}{n}\sum_{i=1}^n (|Z_i|^p-\mathbb{E}[|Z_i|^p])\geq \delta \right]    \label{use_mean_gg}
\end{IEEEeqnarray}
where \eqref{use_inequality_p_geq_1} holds by \eqref{choice_nu_gg} and \eqref{in2};  \eqref{use_mean_gg} holds because  $\mathbb{E}[|Z|^p]=2\sigma^p/p$, see \cite[Eq (3)]{generalized_gaussian_distributions_short}.

By the weak law of large numbers, the right-hand side of \eqref{use_mean_gg} vanishes with $n$, which establishes \eqref{eq:Dn}.

%Since conditions \eqref{eq:Dn} and \eqref{eq:ratio_lim} are 
\section{Proof of Theorem~\ref{theorem_dmc}}\label{sec:proof2}

Define 
\begin{equation} 
\gamma \triangleq \min_{y, x_1,x_2,\tilde{x}_1,\tilde{x}_2} \frac{ P_{Y|X_1,X_2}(y|x_1,x_2)}{ P_{Y|X_1,X_2}(y| \tilde{x}_1,\tilde{x}_2)}
\end{equation} and the minimum costs 
\begin{equation}
c_{\min, \ell} \triangleq \min_{x_\ell \in \mathcal{X}_\ell \backslash \{0\}} c_\ell(x_\ell), \quad \ell \in \{1,2\}.
\end{equation} 
Notice  that the maximum Hamming weight, i.e., the number of symbols different from $0$ for any input sequence $x_\ell^n \in\tilde{ \mathcal{X}}_\ell^n$ is upper bounded by
\begin{equation} \label{eq:kmax}
k_{\max,\ell } \triangleq \frac{\Gamma_\ell(n)}{c_{\min,\ell}},
\end{equation} 
and thus the number of positions on which any two pairs of input sequences $(x_1^n,{x}_2^n)$ and  $(\tilde x_1^n,\tilde {x}_2^n)$ differ is at most 
\begin{equation} \label{eq:k'}
\tau_{\max}\triangleq  2 k_{\max,1}+2 k_{\max,2}.
\end{equation}
Set  also 
\begin{equation} 
k\triangleq  \frac{1}{2}\min \left\{ k_{\max,1},k_{\max,2}\right\}
\end{equation}
and notice that for sufficiently large blocklengths $n$ we have $2k < n$ because $\Gamma_1(n)$ and $\Gamma_2(n)$ grow sublinearly in $n$. \\
Notice finally that, as shown for example in \cite[Equations (50)--(52)]{itw_dichotomy},  the sets  $\tilde{\mathcal{X}}^n_\ell$ grow sublinearly in $n$: 
\begin{equation} \label{eq:zr}
\lim_{n\to \infty} \frac{1}{n} \ln | \tilde{\mathcal{X}}^n_\ell|=0. 
\end{equation}

\subsection{Proof of ~\ref{it:th1})}
Only the converse requires proof.  To this end, 
for all pairs $x_1^n, \tilde{x}_1^n \in \tilde{\mathcal{X}}_1^n$ and $x_2^n, \tilde{x}_2^n \in \tilde{\mathcal{X}}_2^n$ and $y^n \in \mathcal{Y}^n$:
\begin{equation} 
\frac{ P_{Y|X_1,X_2}^{\otimes n}(y^n |x_1^n,x_2^n)}{ P_{Y|X_1,X_2}^{\otimes n}(y^n |x_1^n,x_2^n)} \in \left[  \gamma^{ -\tau_{\max}} ,  \gamma^{ \tau_{\max}}\right],
\end{equation} 
and thus 
\begin{equation} 
\lim_{n\to \infty} \frac{1}{n} \ln \frac{ P_{Y|X_1,X_2}^{\otimes n}(y^n |x_1^n,x_2^n)}{ P_{Y|X_1,X_2}^{\otimes n}(y^n |\tilde{x}_1^n,\tilde{x}_2^n)}=0.
\end{equation} 
The converse is thus obtained directly from  Theorem \ref{thm1} by choosing $\mathcal{D}_n=\mathcal{Y}^n$.

\subsection{Proof of \ref{it:th2})}
\underline{\textit{Converse:}} The converse is directly obtained  from \cite{papamarcou92}
%by noting that the inputs $X_1^n$ and $X_2^n$ take values in alphabets of zero-rate, see \eqref{eq:zr}, and
by giving the decision center direct access to inputs $x_1^n$ and $x_2^n$. 
In this enhanced scenario the outputs $Y^n$ become useless and we fall back to a scenario that is equivalent to the zero-rate scenario studied in \cite{papamarcou92} because the number of different input sequences $|\tilde{\mathcal{X}}^n_1|$ and $|\tilde{\mathcal{X}}^n_2|$ is sublinear in $n$, see \eqref{eq:zr}.\\

\underline{\textit{Achievability:}} The achievability is proved based on the following scheme. Fix $\mu>0$ and pick a set of symbols $x_1, x_1', \tilde{x}_1\in\mathcal{X}_1$, $x_2, x_2',\tilde{x}_2\in\mathcal{X}_2$ and $y,\tilde{y}\in\mathcal{Y}$ satisfying the conditions \eqref{eq:full_conditions}. \\

    \noindent \textit{Sensor $1$:}
    \begin{itemize}
        \item During   the first $k$ channel uses, it sends the symbol  $x_1'$ if $u_1^n \in T_{\mu}(P_{U_1})$; otherwise it sends the symbol $x_1$.
        \item During the next  $k$ channel uses, it  sends the  symbol $\tilde{x}_1$. 
        \item During the remaining channel uses, it sends the $0$ symbol. 
    \end{itemize}
     \textit{Sensor $2$:}
    \begin{itemize}
      \item During the first $k$ channel uses, it  sends the  symbol $\tilde{x}_2$.      
        \item During   the next $k$ channel uses, it sends the symbol  $x_2'$ if $u_2^n \in T_{\mu}(P_{U_2})$; otherwise it sends the symbol $x_2$.
            \item During the remaining channel uses, it sends the $0$ symbol. 
    \end{itemize}
     \textit{Decision center}: It declares $\hat{H} = 0$ if the following three conditions are simultaneously satisfied: 
     \begin{itemize} 
     \item the symbol $y$ occurs during the first $k$ channel outputs $Y_1,\ldots, Y_{k}$, 
     \item the symbol  $\tilde{y}$ occurs during channel outputs $Y_{k+1}, \ldots, Y_{2k}$,
     \item $v^n \in T_{\mu}(P_{V})$; 
     \end{itemize}
otherwise it declares $\hat{H} = 1$. \\
\textit{Analysis}: We start by showing that the type-I error probability of this proposed scheme vanishes when $n\rightarrow \infty$. To this end,  notice that
\begin{IEEEeqnarray}{rCl}
1-\alpha_n&=&\mathbb{P}\left[\hat{H}=0 \; \big| \;H=0\right]\\
&=&\mathbb{P}\left[ y\in Y^{k} , \; \tilde{y}\in Y_{k+1}^{2k} ,  V^n \in \mathcal{T}_{\mu}(P_{V}) \; \big| \;H=0\right] \\
&\geq& \mathbb{P}\left[ y\in Y^{k} ,\;  \tilde{y}\in Y_{k+1}^{2k} , \;  V^n \in \mathcal{T}_{\mu}(P_{V}) , \;  U_1^n \in \mathcal{T}_{\mu}(P_{U_1}), \; U_2^n \in \mathcal{T}_{\mu}(P_{U_2}) \; \big |\;H=0\right] 		\\
&=& \mathbb{P}\left[ V^n \in \mathcal{T}_{\mu}(P_{V}) ,  \; U_1^n \in \mathcal{T}_{\mu}(P_{U_1}), \; U_2^n \in \mathcal{T}_{\mu}(P_{U_2}) \; \big| \;H=0\right] \nonumber\\
&& \cdot \mathbb{P}\left[ y\in Y^{k} , \tilde{y}\in Y_{k+1}^{2k}  \; \big| \; U_1^n \in \mathcal{T}_{\mu}(P_{U_1}) ,  \;U_2^n \in \mathcal{T}_{\mu}(P_{U_2}), H=0\right] \\
\label{dmc_achievability_second_case_step_1}
&=& \mathbb{P}\left[ V^n \in \mathcal{T}_{\mu}(P_{V}) ,  U_1^n \in \mathcal{T}_{\mu}(P_{U_1}), \; U_2^n \in \mathcal{T}_{\mu}(P_{U_2}) \; \big| \;H=0\right] \nonumber\\
&& \cdot \mathbb{P}\left[ y\in Y^{k} ,\;  \tilde{y}\in Y_{k+1}^{2k}  \; \big | \;X_1^{2k}=({x'_1}^{k},\tilde{x}_1^{k}) , X_2^{2k}=(\tilde{x}_2^{k},{x'_2}^{k}) \right] \\
\label{dmc_achievability_second_case_step_2}
&=& \mathbb{P}\left[ V^n \in \mathcal{T}_{\mu}(P_{V}) ,  U_1^n \in \mathcal{T}_{\mu}(P_{U_1}), U_2^n \in \mathcal{T}_{\mu}(P_{U_2}) \; \big| \;H=0\right] \nonumber\\
&& \cdot \mathbb{P}\left[y\in Y^{k} \; \big| \;  X_1^{k}={x_1'}^{k} , X_2^{k}=\tilde{x}_2^{k} \right] \nonumber\\
&& \cdot   \mathbb{P}\left[ \tilde{y}\in Y_{k+1}^{2k} \; \big| \; X_{1,k+1}^{2k}=\tilde{x}_1^{k} ,  X_{2,k+1}^{2k}={x_2'}^{k}\right] \\
&= &\mathbb{P}\left[ V^n \in \mathcal{T}_{\mu}(P_{V}) ,  U_1^n \in \mathcal{T}_{\mu}(P_{U_1}), U_2^n \in \mathcal{T}_{\mu}(P_{U_2}) \; \big| \;H=0\right] \nonumber\\
\label{dmc_achievability_second_case_step_3}
&&  \cdot  \left(1- (1-p_{Y|X_1,X_2}(y|x_1',\tilde{x}_2))^{k} \right) \cdot \left( 1- (1-p_{Y|X_1,X_2}(\tilde{y}|\tilde{x}_1,x_2'))^{k}\right)
\end{IEEEeqnarray}
where \eqref{dmc_achievability_second_case_step_1} follows by the design of the coding scheme and because the channel transition law does not depend on the hypothesis; \eqref{dmc_achievability_second_case_step_2} follows because the channel is memoryless. 
Notice further that by the weak law of large numbers, irrespectively of $\mu$,
\begin{IEEEeqnarray}{rCl}
\lim\limits_{n\rightarrow \infty}  \mathbb{P}\left[ V^n \in \mathcal{T}_{\mu}(P_{V}) ,  U_1^n \in \mathcal{T}_{\mu}(P_{U_1}), U_2^n \in \mathcal{T}_{\mu}(P_{U_2}) \; \big| \;H=0\right] =1.
\end{IEEEeqnarray}
Since $1-p_{Y|X_2,X_1}(y|x_1',\tilde{x}_2)$ and $1-p_{Y|X_1,X_2}(\tilde{y}|\tilde{x}_1,x_2')$ lie in the half-open interval $(0,1]$, we can thus conclude by  \eqref{dmc_achievability_second_case_step_3} that  $\lim \alpha_n=0$ as $n\rightarrow \infty$.
Next, we proceed to upper-bound the type-II error probability. To this end, for each $n$, we introduce the set $\mathcal{B}_{n,\mu}(P_{U_1},P_{U_2},P_{V})$ of $n$-types such that
\begin{IEEEeqnarray}{rCl}
\mathcal{B}_{n,\mu}(P_{U_1},P_{U_2},P_{V})= \left\{ \pi_{u_1^n,u_2^n,v^n}\in\mathcal{P}_n(\mathcal{U}_1\times\mathcal{U}_2\times \mathcal{V}) : u_1^n\in\mathcal{T}_{\mu}\left(P_{U_1}\right),u_2^n\in\mathcal{T}_{\mu}\left(P_{U_2}\right),v^n\in\mathcal{T}_{\mu}\left(P_{V}\right)  \right\},
\end{IEEEeqnarray}
%\begin{IEEEeqnarray}{rCl}
%\mathcal{S}_{n,\mu}= \big\{ \pi_{u_1^n,u_2^n,v^n}\in\mathcal{P}_n(\mathcal{U}_1\times\mathcal{U}_2\times \mathcal{V}) \quad &\textnormal{s.t.} \quad & \forall u_1,u_2,v\in\mathcal{U}_1\times\mathcal{U}_2\times\mathcal{V}  \colon \nonumber\\
%&&| \pi_{u_1^n}(u_1) - P_{U_1}(u_1)| \leq \mu, \nonumber \\
%&& | \pi_{u_2^n}(u_2) - P_{U_2}(u_2)| \leq \mu ,\nonumber \\
%& & | \pi_{v^n}(v) - P_V(v)| \leq \mu%
%\big\},
%\end{IEEEeqnarray}
and notice
\begin{IEEEeqnarray}{rCl}
\beta_n&=&\mathbb{P}\left[\hat{H}=0 |H=1\right]\\
%&= & \mathbb{P}\left[ y^*\in Y^{k} , y'\in Y_{k+1}^k,  V^n \in \mathcal{T}_{\mu}(P_{V}), U_1^n \in \mathcal{T}_{\mu}(P_{U_1})\; \big| \;H=1\right]\nonumber\\
%&+&\underbrace{\mathbb{P}\left[ y^*\in Y^{k} , y'\in Y_{k+1}^k,  V^n \in \mathcal{T}_{\mu}(P_{V}), U_1^n \not\in \mathcal{T}_{\mu}(P_{U_1})\; \big| \;H=1\right]}_{=0}\nonumber\\
&= & \mathbb{P}\left[ y\in Y^{k} , \tilde{y}\in Y_{k+1}^{2k},  V^n \in \mathcal{T}_{\mu}(P_{V}), U_1^n \in \mathcal{T}_{\mu}(P_{U_1}), U_2^n \in \mathcal{T}_{\mu}(P_{U_2})\; \big| \;H=1\right]\\
\label{first_case_achievability_beta}
&\leq & \mathbb{P}\left[ V^n \in \mathcal{T}_{\mu}(P_{V}) ,  U_1^n \in \mathcal{T}_{\mu}(P_{U_1}), U_2^n \in \mathcal{T}_{\mu}(P_{U_2}) \; \big| \;H=1\right]  \\
&=& \sum_{\pi_{u_1^n,u_2^n,v^n}\in \mathcal{B}_{n,\mu}(P_{U_1},P_{U_2},P_{V})}\mathbb{P}\left[(V^n, U_1^n, U_2^n)\in\mathcal{T}_n(\pi_{u_1^n,u_2^n,v^n})\; \big| \;H=1\right] \\
\label{dmc_achievability_second_case_step_4}
&\leq& \sum_{\pi_{u_1^n,u_2^n,v^n}\in \mathcal{B}_{n,\mu}(P_{U_1},P_{U_2},P_{V})}2^{-nD\left(\pi_{u_1^n,u_2^n,v^n}\|Q_{U_1,U_2,V}\right)}\\
\label{dmc_achievability_second_case_step_5}
&\leq & (n+1)^{|\mathcal{U}_1||\mathcal{U}_2||\mathcal{V}|} 2^{-n \min D\left( \pi_{u_1^n,u_2^n,v^n}\|Q_{U_1,U_2,V}\right)} 
\end{IEEEeqnarray}
where the minimum is taken over all types $ \pi_{u_1^n,u_2^n,v^n}\in \mathcal{B}_{n,\mu}(P_{U_1},P_{U_2},P_{V})$. Here, \eqref{dmc_achievability_second_case_step_4} holds by \cite[Theorem 11.1.4]{coverthomas06}; and \eqref{dmc_achievability_second_case_step_5} by \cite[Theorem 11.1.1]{coverthomas06}. 

Dropping the restriction on $\pi_{u_1^n,u_2^n,v^n}$ being an $n$-type
%in the definition of sets $\mathcal{S}_{n,\mu}$
and defining the $\mu$-marginal neighborhood of $P_{U_1,U_2,V}$ as
\begin{IEEEeqnarray}{rCl}
\mathcal{B}_\mu(P_{U_1},P_{U_2},P_{V}) = \big\{ \tilde{P}_{U_1,U_2,V}\in\mathcal{M}(\mathcal{U}_1\times\mathcal{U}_2\times \mathcal{V}) \quad &\textnormal{s.t.} \quad & \forall u_1,u_2,v\in\mathcal{U}_1\times\mathcal{U}_2\times\mathcal{V}  \colon \nonumber\\
&&| \tilde{P}_{U_1}(u_1) - P_{U_1}(u_1)| \leq \mu, \nonumber \\
&& | \tilde{P}_{U_2}(u_2) - P_{U_2}(u_2)| \leq \mu ,\nonumber \\
& & | \tilde{P}_V(v) - P_V(v)| \leq \mu% \mathbb{P}\left[(U_1^n, U_2^n, V^n)\in\mathcal{T}_n(\pi_{U_1,U_2,V})
 \big\},
\end{IEEEeqnarray}
we conclude from \eqref{dmc_achievability_second_case_step_5} that 
\begin{IEEEeqnarray}{rCl}
\beta_n&\leq & (n+1)^{|\mathcal{U}_1||\mathcal{U}_2||\mathcal{V}|} 2^{-n \min D\left( \tilde{P}_{U_1,U_2,V}\|Q_{U_1,U_2,V}\right)}
\end{IEEEeqnarray}
where now the minimum is taken over all distributions $\tilde{P}_{U_1,U_2,V}\in\mathcal{B}_\mu(P_{U_1},P_{U_2},P_{V})$. 
Taking first the limit $n\rightarrow \infty$ and  then $\mu\rightarrow 0$ we finally conclude that 
\begin{IEEEeqnarray}{rCl}
\varliminf_{n\rightarrow \infty} -\frac{1}{n} \ln\beta_n \geq \min D\left( \tilde{P}_{U_1,U_2,V}\|Q_{U_1,U_2,V}\right),
\end{IEEEeqnarray}
where the minimum is now over all distributions $ \tilde{P}_{U_1,U_2,V}\in\mathcal{M}(\mathcal{U}_1\times \mathcal{U}_2\times\mathcal{V})$ with marginals satisfying $\tilde{P}_{U_1}=P_{U_1}$, $\tilde{P}_{U_2}=P_{U_2}$ and $\tilde{P}_V=P_V$.\\

\subsection{Proof of \ref{it:th3})}
\begin{figure}[!htbp]
\begin{center}
        \begin{tikzpicture}[
            nodetype1/.style={
                rectangle,
                %rounded corners,
                minimum width=0.7cm,
                minimum height=0.7cm,
                draw=black,
                font=\normalsize
            },
            nodetype2/.style={
                rectangle,
                %rounded corners,
                minimum width=0.55cm,
                minimum height=0.7cm,
                draw=black,
                font=\normalsize
            },
            tip2/.style={-{Stealth[length=1.5mm, width=1.5mm]}},
            every text node part/.style={align=center}
            ]
            \matrix[row sep=0.3cm, column sep=0.38cm, ampersand replacement=\&]{
            \node (U_1) {$U_1^n$}; \&\& \node (encoder1) [nodetype2]   {Sensor $1$\\$f_1'$};
             \& \& (invisible) \& \& \& (invisible) \& \& (invisible)\& \node (V) {$V^n$};\&; \\
            \&  \& (invisible) \& \& (invisible) \& \& 
            \node (W)  [draw, nodetype1, text width=2cm, text centered]  {\begin{tabular}{c} Channel\\$P_{Y|X_1,X_2}$ \end{tabular}}; \&  \& (invisible)\&
            \node (Y){};
            \&
            \node (decoder) [nodetype2] {Decision center\\$g'$}; \&
            \node (guess) {$\hat{H}$};\\
            \node (U_2) {$U_2^n$};  \& \& \node (encoder2) [nodetype2]  {Sensor $2$\\$f_2'$}; \&
            \node (X){}; \\
            };
            
            \draw[->] (U_1) edge[tip2] node [above] {} (encoder1) ;
            \draw[-]  (encoder1.east) edge[tip2] node [above] (X1) {$X_1^n$} (W.mid west) ;
            \draw[->] (W) edge[tip2] node [above] {$(Y^n,X_1^n)$} (decoder) ;
            \draw[->] (decoder) edge[tip2] node [above] {} (guess) ;
            \draw[-] (encoder2.east) edge[tip2] node [below] (X2) {$X_2^n$} (W.base west) ;
            \draw[->] (U_2) edge[tip2] (encoder2) ;
            \draw[->] (V) edge[tip2] (decoder) ;
        \end{tikzpicture}
    \caption{Enhanced distributed hypothesis testing setup where the output consists of the pair $(Y,X_1)$.}
     \label{fig:0}
     \end{center}
\end{figure}
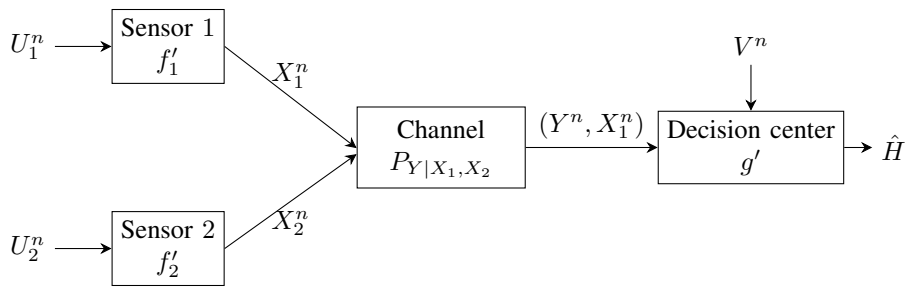

\paragraph{Achievability} Pick $x_1, x_1', \tilde{x}_2, y^*$ satisfying Condition \eqref{definition_C_sparse_full}. \\
 \textit{Sensor $1$}: During the first $k$ channel uses, it sends the symbol  $x_1'$ if $U_1^n \in T_{\mu}(P_{U_1})$; otherwise it sends the symbol $x_1$. Subsequently, it sends the 0 symbol. \\
 \textit{Sensor $2$}:  During the first $k$ channel uses, it sends  the symbol $\tilde{x}_2$. Subsequently, it sends the 0 symbol. \\  
\textit{Decision center}: It declares $\hat{H} = 0$ if the following two conditions are simultaneously satisfied: 
     \begin{itemize} 
     \item the symbol $y^*$ occurs in the output sequence $Y^k$ ,
     \item $v^n \in T_{\mu}(P_{V})$; 
     \end{itemize}
otherwise, it declares $\hat{H} = 1$. \\
In the following, we analyze the type-I and type-II error probabilities of this scheme. First, we show that the type-I error probability of this proposed scheme vanishes using the same proof steps as in \eqref{dmc_achievability_second_case_step_3}:
\begin{IEEEeqnarray}{rCl}
1-\alpha_n%&=&\mathbb{P}\left[\hat{H}=0 |H=0\right]\nonumber\\
%&=&\mathbb{P}\left[ y^*\in Y^k ,  V^n \in \mathcal{T}_{\mu}(P_{V}) \bigg|H=0\right] \nonumber\\
%&\geq& \mathbb{P}\left[ y^*\in Y^k ,  V^n \in \mathcal{T}_{\mu}(P_{V}) ,  U_1^n \in \mathcal{T}_{\mu}(P_{U_1})\; \big| \;H=0\right] 		\nonumber\\
%&=& \mathbb{P}\left[ V^n \in \mathcal{T}_{\mu}(P_{V}) ,  U_1^n \in \mathcal{T}_{\mu}(P_{U_1}) \; \big| \;H=0\right] \nonumber\\
%&& \cdot \mathbb{P}\left[ y^*\in Y^k \; \big| \; U_1^n \in \mathcal{T}_{\mu}(P_{U_1}), H=0\right] \nonumber\\
%&= & \mathbb{P}\left[ V^n \in \mathcal{T}_{\mu}(P_{V}) ,  U_1^n \in \mathcal{T}_{\mu}(P_{U_1}) \; \big| \;H=0\right] \nonumber\\
&\geq& \mathbb{P}\left[ V^n \in \mathcal{T}_{\mu}(P_{V}) ,  U_1^n \in \mathcal{T}_{\mu}(P_{U_1}) \; \big| \;H=0\right] \nonumber\\
\label{dmc_achievability_third_case_step_1}
&&  \cdot  \left(1- (1-p_{Y|X_1,X_2}(y^*|x_1',\tilde{x}_2))^{k} \right).
\end{IEEEeqnarray}
By the weak law of large numbers, irrespectively of $\mu$,
\begin{IEEEeqnarray}{rCl}
\lim\limits_{n\rightarrow \infty}  \mathbb{P}\left[ V^n \in \mathcal{T}_{\mu}(P_{V}) ,  U_1^n \in \mathcal{T}_{\mu}(P_{U_1}) \; \big| \;H=0\right] =1.
\end{IEEEeqnarray}
Since $1- p_{Y|X_1,X_2}(y^*|x_1',\tilde{x}_2)$ lies in the half-open interval $(0,1]$, by \eqref{dmc_achievability_third_case_step_1}, we conclude that $1-\alpha_n$ tends to $1$ as $n\rightarrow \infty$ and consequently $\lim \alpha_n=0$, when $n\rightarrow \infty$.
Next, we analyze the type-II error probability, using the same proof steps as in \eqref{dmc_achievability_second_case_step_5}: 
\begin{IEEEeqnarray}{rCl}
\beta_n%&=&\mathbb{P}\left[\hat{H}=0 |H=1\right]\nonumber\\
%&= & \mathbb{P}\left[ y^*\in Y^k ,  V^n \in \mathcal{T}_{\mu}(P_{V}), U_1^n \in \mathcal{T}_{\mu}(P_{U_1})\; \big| \;H=1\right]\nonumber\\
%&\leq & \mathbb{P}\left[ V^n \in \mathcal{T}_{\mu}(P_{V}) ,  U_1^n \in \mathcal{T}_{\mu}(P_{U_1})\; \big| \;H=1\right]\nonumber\\
%&=&\sum_{\pi_{u_1^n,v^n}\in\mathcal{B}_{n,\mu}(P_{U_1},P_V)}\mathbb{P}\left[(V^n, U_1^n)\in\mathcal{T}_n(\pi_{V,U_1})\; \big| \;H=1\right] \nonumber\\
%\label{dmc_achievability_third_case_step_2}
%&\leq& \sum_{\pi_{u_1^n,v^n}\in\mathcal{B}_{n,\mu}(P_{U_1},P_V)}2^{-nD\left(\pi_{u_1^n,v^n}||Q_{V,U_1}\right)}\\
\label{dmc_achievability_third_case_step_3}
&\leq & (n+1)^{|\mathcal{U}_1||\mathcal{V}|} 2^{-n \min D\left( \pi_{u_1^n,v^n}\|Q_{U_1,V}\right)}
\end{IEEEeqnarray}
where %\eqref{dmc_achievability_third_case_step_2} follows by \cite[Theorem 11.1.4]{coverthomas06};
%\eqref{dmc_achievability_third_case_step_3} by \cite[Theorem 11.1.1]{coverthomas06}
such that the minimum is over all $n$-types in the set
\begin{IEEEeqnarray}{rCl}
\mathcal{B}_{n,\mu}(P_{U_1},P_V)= \left\{ \pi_{u_1^n,v^n}\in\mathcal{P}_n(\mathcal{U}_1\times \mathcal{V}) : u_1^n\in\mathcal{T}_{\mu}\left(P_{U_1}\right),v^n\in\mathcal{T}_{\mu}\left(P_{V}\right)  \right\}.
\end{IEEEeqnarray}
Next, we upper-bound \eqref{dmc_achievability_third_case_step_3}, taking the minimum over the $\mu$-marginal neighborhood of $P_{U_1,V}$
\begin{IEEEeqnarray}{rCl}
\mathcal{B}_{\mu}(P_{U_1},P_V)= \big\{ \tilde{P}_{U_1,V}\in\mathcal{M}(\mathcal{U}_1\times \mathcal{V}) \quad &\textnormal{s.t.} \quad & \forall u_1,v\in\mathcal{U}_1\times\mathcal{V}  \colon \nonumber\\
&&| \tilde{P}_{U_1}(u_1) - P_{U_1}(u_1)| \leq \mu, \nonumber \\
& & | \tilde{P}_V(v) - P_V(v)| \leq \mu
 \big\}.
\end{IEEEeqnarray}
Finally, we let $n\rightarrow \infty$ and $\mu\rightarrow 0$ and we  conclude that 
\begin{IEEEeqnarray}{rCl}
\varliminf_{n\rightarrow \infty} -\frac{1}{n} \ln\beta_n \geq \min D\left( \tilde{P}_{V,U_1}\|Q_{V,U_1}\right),
\end{IEEEeqnarray}
where the minimum is over all distributions $\tilde{P}_{V,U_1}$ with marginals $P_V$ and  $P_{U_1}$.
\paragraph{Converse}
We first introduce some useful notation. 
        Define %This follows by translating Theorem~\ref{thm:Gaussian} and its proof to discrete alphabets. In fact, define 
\begin{equation}
\label{def_gamma_1}
\gamma_1 \triangleq \min_{\substack{y, x_1,x_2,\tilde{x}_1,\tilde{x}_2 \colon \\
P_{Y|X_1,X_2}(y|x_1,x_2)>0}}\frac{ P_{Y|X_1,X_2}(y|x_1,x_2)}{ P_{Y|X_1,X_2}(y| \tilde{x}_1,\tilde{x}_2)}>0
\end{equation} 
We consider an enhanced setup where the detector has access not only to the side information $V^n$ and the channel output $Y^n$, but also directly to the channel input $X_1^n$. The Stein-exponent of this enhanced setup  (depicted in Figure~\ref{fig:0}) is an upper bound to the Stein-exponent of our original setup.  

Encoding functions in the enhanced setup are the same as in our original setup and are denoted $f_1'$ and $f_2'$. The decision function is denoted $g'$, and the corresponding type-I and type-II error probabilities  $\alpha_n'$ and $\beta_n'$. 

\begin{figure}[!htbp]
\begin{center}
        \begin{tikzpicture}[
            nodetype1/.style={
                rectangle,
                %rounded corners,
                minimum width=0.7cm,
                minimum height=0.7cm,
                draw=black,
                font=\normalsize
            },
            nodetype2/.style={
                rectangle,
                %rounded corners,
                minimum width=0.55cm,
                minimum height=0.7cm,
                draw=black,
                font=\normalsize
            },
            tip2/.style={-{Stealth[length=1.5mm, width=1.5mm]}}
            ]
            \matrix[row sep=0.3cm, column sep=0.38cm, ampersand replacement=\&]{
            \& \& \node (V) {$(V^n,X_1^n)$}; \\
            \& \& \node (invisible) {}; \& \\
            \node (W)  [draw, nodetype1, text width=2cm, text centered]  {$P_{\tilde{Y}^n|X_1^n,V^n}$}; \&
            \node (Y_tilde){};
            \&
            \node (decoder) [nodetype2] {$\tilde{g}$}; \&
            \node (guess) {$\hat{H}$};\\
            };
            
            \draw[arrows = {-Latex[length=1pt]}] (W) edge[tip2] node [above] {$\tilde{Y}^n$} (decoder) ;
            \draw[->] (decoder) edge[tip2] node [above] {} (guess) ;
            \draw[-{Stealth[length=1.5mm, width=1.5mm]}] (invisible.center) -| (W) ;
            \draw[-{Stealth[length=1.5mm, width=1.5mm]}] (invisible.center) -- (decoder) ;
            \draw[-] (V) -- (invisible.center) ;
        \end{tikzpicture}
\caption{Randomized  hypothesis test.}

\label{fig:2b}
\end{center}
\end{figure}
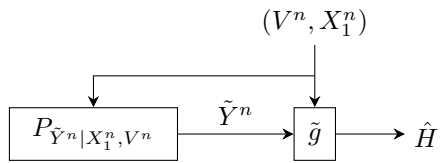

Fix $\epsilon \in [0,1)$ and consider any sequence of encoding and decision functions  $(f_1',f_2',g')$ such that $\varlimsup\alpha_n'\leq\epsilon<1$, $n\rightarrow \infty$. 
For the chosen decision function $g'$ and a fixed blocklength $n$, define for each observation $v^n\in\mathcal{V}^n$: \begin{IEEEeqnarray}{rCl}
\mathcal{A}(v^n) \triangleq \{ (x_1^n,y^n) \in \mathcal{X}_1^n \times \mathcal{Y}^n \colon g( v^n, x_1^n, y^n)=0\}
\end{IEEEeqnarray}
and 
\begin{equation} 
\mathcal{A}'(v^n) \triangleq \{ (x_1^n,y^n) \in \mathcal{X}_1^n \times \mathcal{Y}^n \colon g( v^n, x_1^n, y^n)=0, \;\; P_{Y|X_1,X_2}(y|x_1,x_2) >0 \;\; \forall x_2\in\mathcal{X}_2 \}\footnote{By our assumption, for given $(y,x_1)$  either $P_{Y|X_1,X_2}(y|x_1,x_2) >0$ for all $x_2$ or for no $x_2$.} 
\end{equation} 
Notice that the two regions  $\mathcal{A}(v^n) $ and $\mathcal{A}'(v^n) $ have same probability to occurr under both $H=0$ and $H=1$.
Define further the conditional pmf
\begin{equation} \label{eq:cond2}
P_{\tilde{Y}^n|X_1^n,V^n}(y^n|x_1^n,v^n)\triangleq \mathbb{P}[ Y^n = y^n| X_1^n=x_1,V^n=v^n, H=0],
\end{equation} 
and introduce the random binary hypothesis testing setup where the decision center observes $(X_1^n,V^n)$ which has the same distribution as in our original setup, and has access to the local randomness $\tilde{Y}^n\sim P_{\tilde{Y}^n|X_1^n,V^n}$, irrespectively of the hypothesis $H\in\{0,1\}$. The randomized test is depicted in Figure~\ref{fig:2b}.

We consider the same decision function $g'$ for this auxiliary setup as in the enhanced setup, simply applied to $\tilde{Y}^n$ instead of $Y^n$. The  type-I and type-II error probabilities  of this test are:
\begin{subequations} 
\begin{IEEEeqnarray}{rCl}
\tilde \alpha_n & \triangleq& \mathbb{P}[  g'(X_1^n,V^n,\tilde{Y}^n)=1 | H=0]  \\
\tilde \beta_n & \triangleq  &  \mathbb{P}[  g'(X_1^n,V^n,\tilde{Y}^n)=0 | H=1]  .
\end{IEEEeqnarray}
\end{subequations} 
Notice that the local randomness $\tilde{Y}^n$ follows the same joint distribution  with $(X_1^n,V^n)$ as $Y^n$ under $H=0$. Therefore,  $\tilde{\alpha}_n=\alpha_n'$ and $\varlimsup_{n\to \infty} \tilde{\alpha}_n \leq \epsilon$. Moreover, for the type-II error probabilities of the two tests, we obtain similarly as in the proof of Theorem \ref{thm1}: 
\begin{IEEEeqnarray}{rCl}
    \beta_n'
    &=&\sum_{v^n\in\mathcal{V}^n}\mathbb{P}[(X_1^n,Y^n)\in\mathcal{A}'(v^n),V^n=v^n|H=1]\\
    %&=&\sum_{v^n\in\mathcal{V}^n}\int_{\mathcal{A}(v^n)}T\left(y^k,v^n\right) \dd P_{Y^k,V^n|H}(y^k,v^n|1)\nonumber\\
  %  &=&\sum_{v^n\in\mathcal{V}^n}\int_{\mathcal{A}(v^n)} \dd P_{Y^k,V^n|H}(y^k,v^n|1)\nonumber\\
 %   &= &\sum_{v^n\in\mathcal{V}^n}\left(Q_{V}^{\otimes n}(v^n)\sum_{(x_1^n,y^n)\in \mathcal{A}(v^n)} \sum_{x_2^n} \left( P_{X_1^n,X_2^n|V^n,H}(x_1^n,x_2^n|v^n,1)  P^{\otimes n}_{Y|X_1,X_2}(y^n|x_1^n,x_2^n)\right)\right) \label{eq:e1}\\
    &= &\sum_{v^n\in\mathcal{V}^n}\left(Q_{V}^{\otimes n}(v^n)\sum_{\substack{(x_1^n,y^n)\in \mathcal{A}'(v^n) }} \sum_{x_2^n} \left( P_{X_1^n,X_2^n|V^n,H}(x_1^n,x_2^n|v^n,1)  P^{\otimes n}_{Y|X_1,X_2}(y^n|x_1^n,x_2^n)\right)\right) \\
    &= &\sum_{v^n\in\mathcal{V}^n}\left(Q_{V}^{\otimes n}(v^n)\sum_{\substack{(x_1^n,y^n)\in \mathcal{A}'(v^n) }} P_{X_1^n| V^n,H}(x_1^n,v^n,1) \sum_{x_2^n} \left(P_{ X_2^n|X_1^n,V^n,H}(x_1^n,x_2^n|v^n,1)  P^{\otimes n}_{Y|X_1,X_2}(y^n|x_1^n,x_2^n)\right)\right) \\
    &\geq  & \gamma_1^{ \tau_{\max}} \sum_{v^n\in\mathcal{V}^n}\left(Q_{V}^{\otimes n}(v^n)\sum_{\substack{(x_1^n,y^n)\in \mathcal{A}'(v^n)}} P_{X_1^n| V^n,H}(x_1^n,v^n,1) \sum_{x_2^n} \left(P_{ X_2^n|X_1^n,V^n,H}(x_1^n,x_2^n|v^n,0)  P^{\otimes n}_{Y|X_1,X_2}(y^n|x_1^n,x_2^n)\right)\right) \nonumber \\
    \label{eq:e1d}\\
    &=  & \gamma_1^{ \tau_{\max}} \sum_{v^n\in\mathcal{V}^n}\left(Q_{V}^{\otimes n}(v^n)\sum_{\substack{(x_1^n,y^n)\in \mathcal{A}'(v^n) }} P_{X_1^n| V^n,H}(x_1^n,v^n,1) P_{ Y^n|X_1^n,V^n,H}(x_1^n,x_2^n|v^n,0) \right) \\
    &=  & \gamma_1^{ \tau_{\max}} \sum_{v^n\in\mathcal{V}^n}\left(Q_{V}^{\otimes n}(v^n)\sum_{\substack{(x_1^n,y^n)\in \mathcal{A}'(v^n) }} P_{X_1^n| V^n,H}(x_1^n,v^n,1) P_{ \tilde{Y}^n|X_1^n,V^n}(x_1^n,x_2^n|v^n) \right) \label{eq:e1f}\\
   %    &\geq& \zeta \sum_{v^n\in\mathcal{V}^n}\left(Q_{V}^{\otimes n}(v^n)\int_{\mathcal{A}(v^n) \cap \mathcal{D}_{n}}   \xi(y^n)  \dd P_{Y^n|V^n,H}(y^n|v^n,0)\right)    \label{eq:proof_lemma_1_beta_step_3a}\\
%    &= &  e^{-4\sqrt{kP(n\sigma^2+kP+\delta)} -\frac{2kP}{\sigma^2} } \cdot \sum_{v^n\in\mathcal{V}^n}\left(Q_{V}^{\otimes n}(v^n) \mathbb{P}\left[\tilde{Y}^n\in\left(\mathcal{A}(v^n) \cap \mathcal{D}_{\delta,n}\right)|V^n=v^n, H=1\right] \right)\nonumber\\
%    \label{identifying_beta_tilde}
%    &=&e^{-4\sqrt{kP(n\sigma^2+kP+\delta)}   -\frac{2kP}{\sigma^2} }
& = &\gamma_1^{ \tau_{\max}}  \cdot \tilde{\beta}_n,   \label{identifying_beta_tilde}
\end{IEEEeqnarray}
where $\tau_{\max}$ is defined in \eqref{eq:k'} and grows sublinearly in $n$. Here, \eqref{eq:e1d} holds because the two pairs $(x_1^n,x_2^n)$ and $(\tilde{x}_1^n,\tilde{x}_2^n)$ differ in at most $\tau_{\max}$ positions and by the definition of $\gamma_1$ in \eqref{def_gamma_1};  \eqref{eq:e1f}  holds by the definition of $\tilde{Y}^n$, see \eqref{eq:cond2}. 

Since $\gamma_1$ is a constant and $\tau_{\max}$ grows sublinearly in $n$, see  \eqref{eq:kmax}, \eqref{eq:k'}, and \eqref{eq:stringent}, we conclude from  \eqref{identifying_beta_tilde}  that 
\begin{equation} \label{eq:b}
\varlimsup_{n\to \infty} - \frac{1}{n}\ln  \beta_n' \leq \varlimsup_{n\to \infty} - \frac{1}{n}\ln \tilde{\beta}_n
\end{equation} 
and the Stein-exponent of our original setup cannot be larger than the Stein-exponent of the auxiliary setup where the local randomness $\tilde{Y}^n$ replaces the observation $Y^n$. Consider now the special case of Proposition~\ref{lemma2} where the MAC is the channel $Y=X_1$. The randomized hypothesis test in our enhanced setup is of this form and we can thus deduce that the local randomness $\tilde{Y}^n$ does not increase the Stein-exponent of our enhanced setup.
Without local randomness, the enhanced setup is however equivalent to a single-sensor setup with a noiseless link from Sensor 1 to the decision center, and since the number of input sequences $|\tilde{\mathcal{X}}^n_1|$ is sublinear in $n$, see  \eqref{eq:zr}, the Stein-exponent is upper bounded by the exponent in 
 \cite[Theorem 1]{papamarcou92}, i.e., 
\begin{IEEEeqnarray}{rCl}
\label{dmc_converse_third_case_beta_final}
\varlimsup_{n\to \infty}- \frac{1}{n} \ln\beta_n' \leq \min D\left(\tilde{P}_{U_1,V} \| Q_{U_1,V}\right),
\end{IEEEeqnarray}
where the minimization is over all probability mass functions $\tilde{P}_{U_1,V}$ with marginals $P_{U_1}$ and $P_V$. 

Combining \eqref{dmc_converse_third_case_beta_final} with \eqref{eq:b} concludes the proof.

\subsection{Proof of \ref{it:th4})}
The proof follows the same steps as the one for \ref{it:th3}) by symmetry.

%\begin{remark}
%    Theorem \ref{theorem_dmc} is reminiscent of \cite{sreekumar_without_side_information},
%    %\cite[Remark 4]{sreejith_hypothesis_noisy_dmc},
%    upon noting that, in the notation therein, we consider $\tau=0$, and $E_c=\infty$ in the cases $\mathcal{C}_{\textnormal{sparse}},\mathcal{C}_{\textnormal{sparse-full}},\mathcal{C}_{\textnormal{full-sparse}}$.
%\end{remark}

\section{Summary and Discussion}
\label{section_conclusion}

We characterized the Stein-exponent for two-sensors distributed detection over noisy memoryless channels with stringent input cost constraints that grow sublinearly in the blocklength $n$. For a large class of MACS, like Gaussian MACs and fully-connected DMMACs, the sublinear cost constraints render communication from the two sensors to the decision center useless in terms of Stein's exponent. In these setups, the Stein-exponent coincides with the exponent in the non-distributed case where the decision center has to take its decision solely based on its own observations.  In contrast, for the class of partially-connected DMMACs where certain outputs can be induced only by a subset of the inputs from each user, the Stein-exponent coincides with the exponent in a scenario where communication from both sensors takes place over noiseless links of zero-rate, a scenario solved in \cite{papamarcou92}. For the case where the partial-connectivity only holds from the first sensor but not the second, the Stein-exponent of our setup coincides with the Stein-exponent in a setup without the second sensor and noise-free zero-rate communication from the first sensor.

While this manuscript focuses on two-sensor setups, our proofs and results readily extend to scenarios with an arbitrary number of sensors. 

Comparing our results to the Stein-exponent of distributed hypothesis testing over DMMACs without cost constraints studied in \cite{Michele_noisy_and_MAC}, we observe that the stringent resource constraint severely degrades the Stein-exponent. In particular, without sublinear cost constraints, the Stein-exponent does not degrade to the exponent of the local setup, but the information from the sensor is useful even when communicated over a noisy channel. 

While in this paper we solved the problem for general DMMACs and general cost constraints, we only considered the
class of generalized Gaussian channels with moment constraints. It will be interesting to extend our study to more general classes of
continuous-valued channels.

\appendices
\section{Proof of Proposition~\ref{lemma2}}\label{app:Randomness}

Before proving the proposition, we make the considered setup more precise. 
For each blocklength $n$, let  $\Sn$ denote the local randomness at the decision center that under both hypotheses follows the same distribution
\begin{equation} 
P_{\Sn | V^n}(\sn |v^n). 
\end{equation} 
The decision center thus can now base its decision on the triple $V^n,Y^n,\Sn$, where $V^n,Y^n$ are described as in our original setup and $\Sn$ is the newly introduced local randomness. The decision rule thus is of the form $\tilde{g}(V^n,Y^n,\Sn)$ and $\tilde{\alpha}_n$ and $\tilde \beta_n$  denote the type-I and type-II error probabilities of the  randomized decision rule: 
\begin{IEEEeqnarray}{rCl}
\tilde    \alpha_n&=&\mathbb{P}\left[\tilde g(V^n,Y^n,\Sn)=1\; \big| \;H=0\right]\\
   \tilde \beta_n&=&\mathbb{P}\left[\tilde g(V^n,Y^n,\Sn)=0\; \big| \;H=1\right].
\end{IEEEeqnarray}

To prove the remark, we show that for any choice of the sequence of randomized decision function $\tilde g$ there exists a sequence of deterministic tests $g$ (without local randomness $\Sn$) that achieves same asymptotic error probabilities. 

To this end, choose a sequence $\gamma_n$ satisfying
\begin{subequations} \label{eq:gm}
\begin{IEEEeqnarray}{rCl} 
\lim_{n\to \infty} \gamma_n &= & 0\\
\lim_{n\to \infty} \frac{1}{n} \ln \gamma_n &= & 0. 
\end{IEEEeqnarray} 
\end{subequations}

For each blocklength $n$,  define a new deterministic decision rule
\begin{equation} 
{g}(v^n, y^n) = \mathbbm{1}\left\{  \Pr\left[ \tilde g\left( v^n, y^n, \Sn\right) =0 \; \right] \leq \gamma_n  \right\}
\end{equation}
and the associated 
acceptance region 
\begin{equation} \label{eq:Gn}
\mathcal{G}_{\gamma_n} \triangleq \left\{ (v^n,y^n) \colon  \Pr\left[ \tilde{g}\left( v^n, {y}^n, \Sn\right) =0   \right] > \gamma_n  \right\} . 
\end{equation}

Using the definition in \eqref{eq:Gn}, we can relate the error probabilities of the two tests as follows:
\begin{IEEEeqnarray}{rCl}
1- \tilde \alpha_n& = & \sum_{(v^n,y^n)} \Big(\Pr[V^n=v^n,Y^n=y^n|H=0] \Pr\left[ \tilde g( v^n, {Y}^n,\Sn) = 0  \;  \Big | \;  V^n=v^n,Y^n=y^n \right]  \Big)\\
& \leq & \sum_{(v^n,y^n)\in \mathcal{G}_{\gamma_n}}  \Pr[V^n=v^n,Y^n=y^n|H=0]\nonumber \\
&& +  \sum_{(v^n,y^n)\notin \mathcal{G}_{\gamma_n}}\Big(\Pr[V^n=v^n,Y^n=y^n|H=0]\underbrace{ \Pr\left[ \tilde g( v^n, {y}^n,\Sn) =0  \;  \Big | \;  V^n=v^n,Y^n=y^n \right ]}_{\leq \gamma_n} \Big) \\
& \leq & 1- {\alpha}_n + \gamma_n\label{eq:at}
\end{IEEEeqnarray}
and as
\begin{IEEEeqnarray}{rCl}
\tilde \beta_n& = & \sum_{(v^n,y^n)} \Pr[V^n=v^n,Y^n=y^n|H=1] \Pr\left[ \tilde{g}( v^n, {y}^n, \Sn) =0  \;  \Big | \; V^n =v^n,Y^n=y^n\right] \\
& \geq &  \sum_{(v^n,y^n) \in \mathcal{G}_{\gamma_n}}  \Big(\Pr[V^n=v^n,Y^n=y^n|H=1] \underbrace{ \Pr\left[ \tilde{g}( v^n, {y}^n, \Sn)  =0  \;  \Big | \; V^n =v^n,Y^n=y^n \right ]}_{> \gamma_n} \Big) \\
\label{beta_bound_randomized_test}
& \geq & {\beta}_n \cdot \gamma_n.\label{eq:bt}
\end{IEEEeqnarray}

Combining \eqref{eq:at} and \eqref{eq:bt} with   \eqref{eq:gm}, we conclude 
\begin{IEEEeqnarray}{rCl} 
\varlimsup_{n\to \infty}  \alpha_n &\leq  &\varlimsup_{n\to \infty}\tilde  \alpha_n \\
\varlimsup_{n\to \infty} - \frac{1}{n} \ln  \beta_n &\geq &  \varlimsup_{n\to \infty} - \frac{1}{n} \ln\tilde  \beta_n.
\end{IEEEeqnarray}
This establishes that the Stein-exponent without randomized decision needs to be at least as large as the Stein-exponent with randomized decisions.

%\bibliographystyle{IEEEtran}
%\begin{small}
%\bibliography{biblio}
%\end{small}

\begin{small}

\end{small}

\end{document}